# Second-generation stoichiometric mathematical model to predict methane emissions from oil sands tailings

Jude D. Kong[1,2], Hao Wang[2*†], Tariq Siddique[3*‡], Julia Foght[4], Kathleen Semple[4], Zvonko Burkus[5], and Mark A. Lewis[2,4]

[1]Center for Discrete Mathematics and Theoretical computer Science, 96 Frelinghuysen Road Piscataway, NJ 08854-8018, USA
[2]Department of Mathematical and Statistical Sciences, University of Alberta, Edmonton, AB T6G 2G1, Canada
[3]Department of Renewable Resources, University of Alberta, Edmonton, AB T6G 2G7, Canada
[4]Department of Biological Sciences, University of Alberta, Edmonton, AB T6G 2E9, Canada
[5]Alberta Environment and Parks, Government of Alberta, Edmonton, Canada

Corresponding authors' emails:
*† Mathematical approach (Hao Wang); hao8@ualberta.ca
*‡ Biological approach (Tariq Siddique); tariq.siddique@ualberta.ca

## ABSTRACT

Microbial metabolism of fugitive hydrocarbons produces greenhouse gas (GHG) emissions from oil sands tailings ponds (OSTP) and end pit lakes (EPL) that retain semisolid wastes from surface mining of oil sands ores. Predicting GHG production, particularly methane ($CH_4$), would help oil sands operators mitigate tailings emissions and would assist regulators evaluating the trajectory of reclamation scenarios. Using empirical datasets from laboratory incubation of OSTP sediments with pertinent hydrocarbons, we developed a stoichiometric model for $CH_4$ generation by indigenous microbes. This model improved on previous first-approximation models by considering long-term biodegradation kinetics for 18 relevant hydrocarbons from three different oil sands operations, lag times, nutrient limitations, and microbial growth and death rates.

Laboratory measurements were used to estimate model parameter values and to validate the new model. Goodness of fit analysis showed that the stoichiometric model predicted $CH_4$ production well; normalized mean square error analysis revealed that it surpassed previous models. Comparison of model predictions with field measurements of $CH_4$ emissions further validated the new model. Importantly, the model also identified parameters that are currently lacking but are needed to enable future robust modeling of $CH_4$ production from OSTP and EPL in situ.

**GRAPHICAL ABSTRACT**

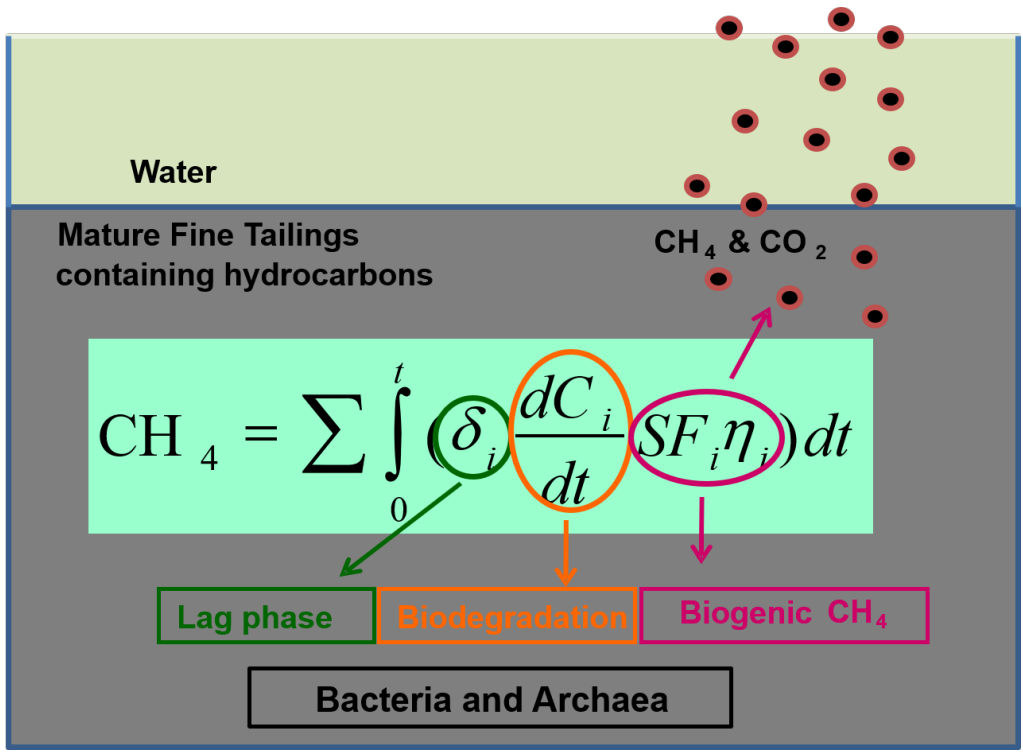

Water

Mature Fine Tailings
containing hydrocarbons

$CH_4$ & $CO_2$

$$CH_4 = \sum \int_0^t (\delta_i \frac{dC_i}{dt} SF_i \eta_i) dt$$

Lag phase    Biodegradation    Biogenic $CH_4$

**Bacteria and Archaea**

## 1. Introduction

Alberta's oil sands ("tar sands") industry is a major economic driver in Canada, currently producing ~2.3 million barrels oil $d^{-1}$ and expected to reach 4 million barrels $d^{-1}$ by 2024 (Government of Alberta, 2019a). However, the oil sands sector has come under international scrutiny regarding GHG emissions and other environmental issues. Oil sands mining and upgrading were responsible for ~24.5% of Alberta's overall GHG emissions in 2012, of which $CH_4$ represented ~6% of oil sands emissions (Alberta Greenhouse Gas Report, 2016). In addition to these production operations, the storage and management of aqueous slurries of surface-mined ore processing wastes in OSTP (Figure S1) contributes substantially to $CH_4$ and carbon dioxide ($CO_2$) emissions (Burkus et al., 2014; Siddique et al., 2008). Total fugitive GHG emissions from major oil sands operators' OSTP, measured *in situ* using floating flux chambers in 2011 were calculated to be 2.8 million tonnes $CO_2$ equivalent per year (Burkus et al., 2014)*.* Furthermore, proposed implementation of EPL as a long-term reclamation strategy for OSTP sediments (Figure S1) may contribute additional GHG emissions for an unknown timespan.

Regulations require that, with few exceptions, oil sands tailings and process waters be retained on-site. During five decades of retention enormous volumes of tailings have accumulated, estimated at 1.21 billion $m^3$ in 2016 (Government of Alberta, 2019b). As the fluid tailings in OSTP age, the suspended clay fines settle via several mechanisms (Siddique et al., 2014) to become anaerobic mature fine tailings (MFT) having a solids content >30 wt% and possessing both an active microbiota and residual diluent in progressive stages of selective biodegradation (Fig S2 in Foght et al., 2017). The depth of OSTP is typically >30 m, essentially isolating the aged lower strata of dense water-saturated sediment from the younger semi-fluid surface strata and overlying water during burial over time. The use of EPL has been proposed to reintegrate the accumulated tailings into the on-site environment (Charette et al., 2012). In this reclamation scenario, after years or decades of residence in OSTP, MFT would be transported to mined-out pits and capped with fresh water and/or process-affected water. This is intended to establish a sustainable aquatic system (i.e., an EPL) that, with time, should support economic, ecological and/or societal uses (Charette et al., 2012). However, ebullition of GHG from underlying sediments may delay EPL ecosystem development by dispersing fine sediments into the overlying water layer along, potentially co-transporting some constituents of concern. Thus, GHG emissions from oil sands tailings repositories are problematic from global warming as well as ecological standpoints.

GHG emissions from OSTP and EPL result primarily from anaerobic biodegradation of diluent hydrocarbons (naphtha or light paraffins) introduced into tailings during aqueous extraction of bitumen from oil sands ore (Figure S1; reviewed in Foght et al., 2017)) The diluents, specific to each operator, facilitate separation of bitumen from water and mineral solid particles during froth treatment and reduce bitumen viscosity in preparation for processing and/or transport. Most of the diluent is recovered from the froth treatment tailings for re-use, but a small proportion remains in the tailings slurry that comprises alkaline water, sand, silt, clays and unrecovered bitumen. These fresh tailings, as well as other tailings streams that have not been exposed to diluent, are deposited in OSTP where indigenous anaerobic microbial communities biodegrade the labile hydrocarbons to $CH_4$ and $CO_2$ (Abu Laban et al., 2015; Penner and Foght, 2010; Mohamad Shahimin et al., 2016; Siddique et al., 2011). Although naphtha and paraffinic diluents are considered the major carbon sources for microbes in OSTP (Foght et al., 2017), only certain of their hydrocarbon components are known to be labile (biodegradable) under anaerobic conditions, whereas others are recalcitrant (slowly or incompletely biodegraded) or are completely resistant to biodegradation (Siddique et al., 2018). Although bitumen is the overwhelming organic constituent of fresh tailings, it predominantly comprises recalcitrant hydrocarbons: only a small proportion may be labile and the contribution of bitumen to biogenic GHG is thought to be negligible in proportion to that of diluent (Foght et al., 2017).

The importance of modeling GHG emissions is clear to regulators and oil sands operators, as it provides a rationale for mitigating GHG mitigation efforts and managing OSTP and EPL. However, field data (e.g., concentrations of individual hydrocarbons in OSTP, nutrient concentrations, biomass) needed for modeling are generally unavailable either because collection of such data is technologically difficult or because key model parameters have not previously been identified as necessary. Therefore, we have cultivated MFT in laboratory cultures analogous to OSTP and EPL for use in initial modeling efforts. A previous study (Siddique et al., 2008) used limited data available from short-term (<1 yr) laboratory studies measuring biodegradation of a small subset of components (Siddique et al., 2007, 2006) in a single naphtha diluent to develop zero- and first-order kinetic models for estimating $CH_4$ production potential from a single OSTP. That first approximation model predicted in situ $CH_4$ production volumes reasonably consistent with emissions measured in situ (Siddique et al., 2008). However, in the decade since that work, additional components of naphtha and paraffinic diluent have been shown to support methanogenesis from MFT during extended laboratory incubation (up to 6.5 y; Abu Laban et al., 2015; Mohamad Shahimin  et al., 2016; Siddique et al., 2015, 2011). This finding increases theoretical GHG emissions, especially from recalcitrant hydrocarbons previously not considered in the previous model and over extended time scales more relevant to long-term retention of tailings. Additionally, data are now available for additional OSTP receiving different diluents and therefore having unique microbial communities (Wilson et al., 2016) with different $CH_4$ production potentials, and the effect of potentially growth-limiting nutrients in situ such as nitrogen has begun to be examined (Collins et al., 2016). Also, the first EPL field trial recently was established and $CH_4$ has been detected in surface layers (Risacher et al., 2018). The greatly expanded data set and a broader understanding of oil sands tailings microbiology (Foght et al., 2017) enable and have driven development of the improved and flexible model for $CH_4$ generation described here.

The goals of the new stoichiometric model were: (1) to expand $CH_4$ predictive capability by considering methanogenic biodegradation of a wider range of hydrocarbons only recently shown to be labile over longer incubation times; (2) for the first time to consider OSTP that receive diluents having different compositions and that harbour different microbial communities; (3) to account for the effects of nutrient limitation on $CH_4$ generation, particularly available nitrogen; (4) to compare model predictions with field measurements of $CH_4$ emissions to validate the model and reveal any shortcomings; (5) to consider differences in GHG emission trajectories between OSTP and EPL; and (6) to identify parameters essential for future development of a model to predict $CH_4$ emissions in situ in OSTP and EPL.

## 2. Materials and Methods

Although the gaseous products of methanogenic hydrocarbon biodegradation are $CH_4$ and $CO_2$ (Figure S2), the stoichiometric model developed here considers only $CH_4$ production for two reasons: $CH_4$ has a greater greenhouse effect than $CO_2$; and measurement of emissions of $CO_2$ emissions produced in MFT is confounded by abiotic (carbonate dissolution) and biogeochemical (mineral precipitation and dissolution) interactions with tailings minerals (Siddique et al., 2014), complicating measurement and modeling.

Methane production from hydrocarbons involves two microbial processes: the oxidation of labile hydrocarbons to simple organic compounds by Bacteria and the conversion of those compounds to $CH_4$ and $CO_2$ by Archaea (Figure S2). Therefore, the model was developed in two modules. The first module (section 2.1) comprising two systems of equations describes bacterial biodegradation of 18 hydrocarbon substrates (see section 2.3.1 for selection rationale) and includes formation of microbial biomass. The second module (section 2.2) considers archaeal $CH_4$ generation from bacterial metabolites. Model parameters unavailable in the literature were estimated by data fitting using laboratory measurements (section 2.3). The model then was quantitatively validated by comparison (1) to measurements from independent but analogous laboratory experiments conducted using oil sands tailings incubated with whole diluents or components of naphtha or paraffinic diluents and (2) to field measurements of $CH_4$ emissions from OSTP (section 2.4). Finally the model was qualitatively assessed using phase plane analysis to illustrate $CH_4$ emission trajectories in OSTP and EPL (section 2.5 and Supporting Material section S3). Terms used in model development are defined in Table 1.

## 2.1 Biodegradation and biomass module development.

Direct measurement of hydrocarbon biodegradation kinetics in OSTP and EPL is technically infeasible. Therefore this module describes the dynamics of $CH_4$ production from MFT incubated with cognate naphtha or paraffinic diluents under laboratory conditions analogous to those expected in OSTP or EPL. A brief description of previously published cultivation methods used to generate model data is given in Supplementary Material section S1.

Microbial biomass can change as a result of two competing processes (growth and death). Because hydrocarbon biodegradation is initiated by Bacteria and not by the archaeal methanogens (Figure S2), this module considers only bacterial kinetics. The per cell bacterial growth rate is assumed to follow Liebig's law of the minimum (Sterner and Elser, 2002) stating that growth rate is proportional to the most limiting resource available. The model assumes, based on chemical analysis of oil sands tailings (Collins, 2013; Penner and Foght, 2010) that all relevant nutrients except biologically-available nitrogen (defined in Table 1) and/or labile carbon are present at non-limiting concentrations in OSTP and EPL. Therefore the bacterial growth rate is modeled as a function only of the mass of biologically-available nitrogen ($N_A$) and labile hydrocarbons ($C_i$, the mass of labile hydrocarbons in the system for $i=1\ldots n$, assuming $n$ discrete labile hydrocarbons in the system). Assuming that there is negligible input of $N_A$ with fresh tailings, no outflow of soluble $N_A$ and no loss of gaseous $NO_x$, we take the total nitrogen ($N_T$) in these systems to be constant. With this assumption, the subset of $N_T$ available for bacterial growth ($N_A$) is given by $N_A = N_T - \theta B$ where $\theta$ is the ratio of nitrogen to carbon in the total microbial biomass B, and $\theta$ is assumed to be constant (Makino et al., 2003). The Monod functions $f(N_A) = \frac{N_A}{N_A + K_f}$ and $g(C_i) = \frac{C_i}{C_i + K_{g_i}}$ are used to model the nitrogen- and carbon-dependent growth rates respectively, where $K_f$ is the $N_A$-dependent half-saturation constant; $K_{g_i}$ is the $C_i$-dependent half-saturation constant; and $C_i{}^{in}$ is the inflow of $C_i$ to the system. Thus, the $C_i$-dependent per cell bacterial growth rate µ is given by $\mu_i \min\{f(N_A), g(C_i)\}$, where $\mu_i$ is the maximum growth rate of Bacteria growing on only the hydrocarbon $C_i$ present and is unique for each labile hydrocarbon. Hence the total per cell growth rate of Bacteria is $\sum_{i=1}^{n} \mu_i \min\{f(N_A), g(C_i)\}$.

The biodegradation rate of each labile hydrocarbon $i$ is assumed to be proportional to the bacterial growth rate due to its consumption, i.e., [per cell bacterial growth rate due to each hydrocarbon] $\propto$ [biodegradation rate of hydrocarbon]. This implies that [the per cell bacterial growth rate supported by each labile hydrocarbon $i$)] = $r_i$[the per cell biodegradation rate of that hydrocarbon] where $r_i$ is a proportionality constant reflecting the efficiency of bacterial conversion of substrate into biomass. Hence, [the per cell biodegradation rate of each labile hydrocarbon] = $\frac{1}{r_i}$ [the per cell bacterial growth rate supported by labile hydrocarbons], i.e., [the per cell biodegradation rate of each hydrocarbon] = $\sum_{i=1}^{n} \frac{1}{r_i} \mu_i \min\{f(N_A), g(C_i)\}$. Archaeal growth and death are considered in the second module (section 2.2).

We assume that microbial death rate ($d$) is constant in the laboratory cultures and that nutrients in dead microbial biomass are quickly recycled back into labile carbon and nitrogen ($N_A$). The fraction of $C_i$ recycled from dead biomass $b$ is assumed to be a constant $\beta_i$ where $0 < \beta_i < 1$.

In accordance with laboratory observations (Mohamad Shahimin and Siddique, 2017a, 2017b, Siddique et al., 2007, 2006), the model assumes that onset of biodegradation of each hydrocarbon begins after a unique lag period, $\lambda_i$. The above assumptions lead to the following system of equations:

$$g(C_i) = \begin{cases} 0, & t < \lambda_i \\ \dfrac{C_i}{K_{g_i} + C_i}, & t \geq \lambda_i \end{cases}$$

$$\frac{dB}{dt} = B \sum_{i=1}^{n} \mu_i \min\left\{\frac{N_A}{K_f + N_A}, g(C_i)\right\} - dB, \qquad (1)$$

$$\frac{dC_i}{dt} = \frac{-1}{r_i} \mu_i B \min\left\{\frac{N_A}{K_f + N_A}, g(C_i)\right\} + \beta_i dB + C_i^{in},$$

$$N_A = N_T - \theta B,$$

$$B(0) > 0, C_i(0) \geq 0.$$

Since the carbon- and nutrient-dependent growth efficiency parameters describe the main differences in bacterial utilization of different hydrocarbon, the model assumes that parameters such as carbon conversion efficiency, intrinsic bacterial growth rate, and carbon recycling from dead bacteria (negligible in our data fitting), are equivalent for different hydrocarbons; i.e., $\mu_i = \mu$, $r_i = r$, and $\beta_i = \beta$. With this assumption, the system of equations becomes:

$$g(C_i) = \begin{cases} 0, & t < \lambda_i \\ \dfrac{C_i}{K_{g_i} + C_i}, & t \geq \lambda_i \end{cases}$$

$$\frac{dB}{dt} = B \sum_{i=1}^{n} \mu\, min\left\{ \frac{N_A}{K_f + N_A}, g(C_i) \right\} - dB, \qquad (2)$$

$$\frac{dC_i}{dt} = \frac{-1}{r} \mu B\, min\left\{ \frac{N_A}{K_f + N_A}, g(C_i) \right\} + \beta dB + C_i^{in},$$

$$N_A = N_T - \theta B,$$

$$B(0) > 0, C_i(0) \geq 0.$$

To analyze the types of solutions that this model could produce, a steady state analysis was performed. The algebraic analysis is described in Supplementary Material section S2 and is of particular use because it allows solutions to be classified by parameter values.

## 2.2 Methane biogenesis module development

From the preceding biodegradation module, bacterial biodegradation of a hydrocarbon substrate ($C_i$) per unit time yields $\frac{1}{r} \mu B\, min\left\{ \frac{N_A}{K_f + N_A}, g(C_i) \right\}$ units of metabolite(s) corresponding to $C_i$. The metabolite(s) ultimately are converted to $CH_4$ and $CO_2$ ($G_i$) by methanogens (Figure S2). Because methanogens have a slow growth rate compared to that of the hydrocarbon-degrading

Bacteria (being dependent on their metabolism), we assume that the biomass of methanogens in the system is constant. With these additions, the system of equations (2) becomes:

$$(C_i) = \begin{cases} 0, & t < \lambda_i \\ \dfrac{C_i}{K_{g_i} + C_i}, & t \geq \lambda_i \end{cases}$$

$$\frac{dB}{dt} = B \sum_{i=1}^{n} \mu \, min\left\{\frac{N_A}{K_f + N_A}, g(C_i)\right\} - dB \ , \tag{3}$$

$$\frac{dC_i}{dt} = \frac{-1}{r} \mu B \, min\left\{\frac{N_A}{K_f + N_A}, g(C_i)\right\} + \beta dB + C_i^{in} \ ,$$

$$\frac{dG_i}{dt} = \frac{1}{r} \mu B \, min\left\{\frac{N_A}{K_f + N_A}, g(C_i)\right\},$$

$$CH_4 = \sum_{i=1}^{n} \eta_i \Gamma_i G_i,$$

$$N_A = N_T - \theta B,$$

$$B(0) > 0, C_i \geq 0, G_i(0) = 0$$

where, $\Gamma_i$ is the maximum theoretical yield of $CH_4$ expected from biodegradation of one mole of $C_i$. This value can be calculated from Equation (4) (derived from Symons and Buswell, 1933, as implemented by Roberts, 2002) that describes the complete oxidation of hydrocarbons to $CH_4$ and $CO_2$ under methanogenic conditions, namely:

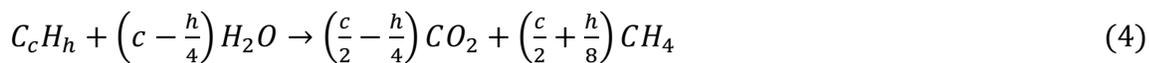

$$C_c H_h + \left(c - \frac{h}{4}\right) H_2 O \rightarrow \left(\frac{c}{2} - \frac{h}{4}\right) CO_2 + \left(\frac{c}{2} + \frac{h}{8}\right) CH_4 \tag{4}$$

where $c$ and $h$ are, respectively, the numbers of carbon and hydrogen atoms in a $C_i$ molecule.

From equation (4), $\Gamma_i = \left(\frac{c}{2} + \frac{h}{8}\right)$. Furthermore, $\eta_i$ is the fraction of the theoretical $CH_4$ yield from the biodegradation of a mole of $C_i$ (i.e., a conversion efficiency factor) and is assumed to be the same for all $C_i$, i.e., $\eta_i = \eta$, with $0 < \eta_i < 1$. The values of $\eta_i$ used in numerical simulations were obtained from (Mohamad Shahimin et al., 2016; Mohamad Shahimin and Siddique, 2017a, 2017b, Siddique et al., 2007, 2006) and Table S1.

## 2.3 Acquisition of laboratory data, parameter estimation and model validation

Our approach was to select a suite of 18 relevant labile hydrocarbons to generate model predictions, then estimate missing model parameters using empirical biodegradation kinetics and $CH_4$ measurements for these hydrocarbons, and finally to test the stoichiometric model quantitatively using measurements from an independent set of laboratory experiments.

### 2.3.1 Model hydrocarbon selection and testing

Fugitive diluent in froth treatment tailings (Fig. S1) is the predominant substrate for methanogenesis in OSTP (Foght et al., 2017). The most commonly used diluents are naphtha and paraffinic solvent. Syncrude Canada Ltd. (Syncrude) and Canadian Natural Resources Ltd. (CNRL) use naphtha, the composition of which differs slightly for each company but which comprises primarily paraffinic (*n*-, *iso*- and *cyclo*-alkanes) and monoaromatic hydrocarbons (predominantly toluene and three xylene isomers), typically in the $C_6$-$C_{10}$ range (Siddique et al., 2008). Canadian Natural Upgrading Limited (CNUL; formerly Shell Albian) uses a paraffinic diluent comprising *n*- and *iso*-alkanes primarily in the $C_5$-$C_6$ range (Mohamad Shahimin and Siddique, 2017a). Published results from laboratory experiments incubating these whole diluents or their major constituents with MFT from Syncrude, CNUL or CNRL (Mohamad Shahimin et al., 2016; Mohamad Shahimin and Siddique, 2017a, 2017b, Siddique et al., 2007, 2006; and Table S1) revealed complete or significant biodegradation of 18 hydrocarbons under methanogenic conditions, including the *n*-alkanes *n*-pentane ($C_5$), *n*-hexane ($C_6$), *n*-heptane ($C_7$),

*n*-octane (C$_8$), *n*-nonane (C$_9$), and *n*-decane (C$_{10}$); the *iso*-alkanes 2-methylpentane (2-MC$_5$), 2-methylhexane (2-MC$_6$), 3-methylhexane (3-MC$_6$), 2-methylheptane (2-MC$_7$), 4-methylheptane (4-MC$_7$), 2-methyloctane (2-MC$_8$), 3-methyloctane (3-MC$_8$) and 2-methylnonane (2-MC$_9$); and the monoaromatics toluene, *o*-xylene and *m*- plus *p*-xylenes (the latter two are not resolved by our gas chromatography column and are therefore reported as a sum). Table 2 lists the 18 labile hydrocarbons selected for model development, the source of biodegradation data, the type of tailings used to generate the data and the parameters estimated using those data.

### 2.3.2 Parameter estimation

The values of many model parameters in the system of equations (3) are not available in the literature, including the initial microbial biomass in OSTP and EPL (B(0)), the nitrogen half-saturation constant (K$_f$), the half-saturation constants of the biodegradable hydrocarbons (K$_{gi}$) and λ$_i$. Because these parameters are related to the biodegradation module, we fit the biodegradation module (system of equations (2)) to data obtained from laboratory biodegradation studies cited above. To estimate these values, we used the nonlinear regression function *nlinfit(.)* in MATLAB, which uses the Levenberg-Marquardt algorithm (Moré, 1978), to fit the solution of the biodegradation module to the data. We provided the function with empirical data (see Table 2 for sources), the time points at which the data were collected (*X*), our simulated results at *X*, and a random initial guess of parameter values. The system was integrated by calling a function that takes as input the initial parameter values, the time at which the empirical data were collected, and for any given time *X* uses the MATLAB function *ode15s(.)* to perform the integration. The solution of the system obtained from the function was then evaluated at *X*, using the MATLAB function *deval(.)*. We also estimated the 95% confidence intervals of the predicted values by using the MATLAB function *nlparci(.)*. To achieve this, we provided this function with the coefficient estimates, residuals and the estimated coefficient covariance matrix from *nlinfit(.)*. Some of the microbial model parameters used in the simulation, namely $\mu$, $r$, and $\theta$, were taken from the literature: the units, values and source of these parameters are provided in Table S2. We assume here that no microbes died during laboratory incubation; thus, in fitting the data to our model, we take $d$ to be zero.

*2.3.3 Model validation against laboratory data*

The new stoichiometric model was then validated against $CH_4$ production data generated in independent but parallel laboratory studies that measured biodegradation of paraffinic diluent in CNUL MFT (Mohamad Shahimin and Siddique, 2017a) and naphtha in Syncrude (Table S1) and CNRL MFT ( Mohamad Shahimin and Siddique, 2017b). To this end, the concentrations of the labile hydrocarbons initially present in each diluent were used in the model to predict $CH_4$ production (Table S7). These predictions were compared with measured $CH_4$ produced by those tailings in independent laboratory experiments using the *goodnessOfFit(.)* function in MATLAB. As input, we provided this function with our test data, the simulated data from our model, and a cost function that determines the goodness of fit. We used the Normalized Mean Square Error (NMSE) function for this statistic, computed as

$$\text{NMSE} = 1 - \frac{\|[\text{actual}] - [\text{predicted}]\|^2}{\|[\text{actual}] - [\text{mean of actual}]\|^2},$$

where $\|.\|$ indicates the 2-norm of a vector, *predicted* is the output simulated by our model, *actual* is the input test data and *mean of actual* is the mean of the test data. $\text{NMSE} \in [-\infty, 1]$ where $-\infty$ indicates a bad fit and 1 a perfect fit.

## 2.4 Quantitative comparison of model prediction and in situ measurement of $CH_4$ emissions from OSTP

To further validate the applicability of model for predicting in situ $CH_4$ emissions, we used (1) a modeling approach where kinetics of $CH_4$ production were estimated to determine the longevity of $CH_4$ emissions, and (2) a direct approach that yielded a ballpark value of potential $CH_4$ emissions. For both approaches we estimated the total mass of diluent entrained in froth treatment tailings entering Syncrude MLSB, CNRL Horizon and CNUL MRM OSTPs in 2016 and 2017 (Table S6) and estimated the mass of individual biodegradable hydrocarbons in diluent (Table S7) using published diluent compositions. To employ the modeling approach, we assumed that these masses of individual hydrocarbons were present at the start of each year (i.e., the model was run as if all the diluent was introduced on January 1 of the year), while acknowledging the continuous input of similar amounts of diluents in the years preceding 2016. Using the estimated parameter values in Table S4, we modeled $CH_4$ production and calculated the predicted cumulative $CH_4$ produced by metabolism of the constituent hydrocarbons over 366 days. The model output was compared with cumulative $CH_4$ emissions measured in flux chambers at the surface of OSTP as reported to the Government of Alberta (unpublished; raw data available upon request) (Table S8). Notably, surface flux measurements of $CH_4$ are not yet available for the single EPL that was established in 2013, so the current comparison is limited to OSTP measurements. In the direct approach, theoretical $CH_4$ production was estimated from the masses of individual hydrocarbons biodegraded to methane using stoichiometric equations as described in Table S8.

## 2.5 Qualitative assessment of model predictions for OSTP and EPL

In addition to quantitative analyses, the model was also qualitatively challenged to predict the trajectories of CH$_4$ generation from OSTP (continuous $C_i^{in}$>0) versus EPL (C$_i$=0) under hypothetical scenarios of carbon or nitrogen availability in situ. Phase plane analysis was performed (Supplemental Material section S3) by assuming that the diluent comprises C$_i$, $i$=1,2,3…,18 are identical and sum up to C$_T$, and that the rate input of all the C$_i$ per unit time into the system is $C_T^{in}$. Equations were solved for microbial biomass versus total carbon content under eight combinations of C$_i$ and N$_A$ limitation over time.

The mathematical model and code are available at http://www.judekong.ca/publication/2019-05-01-Methanebiogenesismodel or from the authors upon request.

## 3. Results and Discussion

Previous zero- and first-order CH$_4$ production models from oil sands tailings (Siddique et al., 2008) used the available limited experimental data for diluent biodegradation and CH$_4$ production from four short-chain $n$-alkanes and four monoaromatic compounds during <1 year incubation with MFT from a single OSTP (Siddique et al., 2007, 2006). Those first approximation models assumed that organic carbon was the sole limiting nutrient in situ and that microbial biomass was constant in OSTP despite receiving continuous and consistent inputs of diluent in froth treatment tailings. The stoichiometric model described here accounts for additional parameters including recently published biodegradation kinetics and CH$_4$ measurements for 18 relevant hydrocarbons including additional $n$-alkanes and, for the first time, $iso$-alkanes, incubated for much longer (up to 6.5 years) with MFT from three different OSTP impacted by distinct diluents. These additional experimental data allow the estimation of some kinetic parameters not previously considered and enable the new model to account for more biological factors than the previous models, so as to be adaptable to future modeling of in situ $CH_4$ production from OSTP and EPL.

## 3.1 Data fitting to biodegradation and methane generation modules.

The biodegradation module was evaluated by fitting system of equations (2) to published experimental data sets for the 18 labile hydrocarbons listed in Table 2. Figures S3-S5 show the simulated biodegradation of diluent *n*-alkanes, monoaromatics and *iso*-alkanes compared with measured biodegradation of these components. We obtained goodness-of-fit statistics (NMSE) ranging from 0.85-1.00 (Table S3). These statistics show that the performance of the module with respect to the training data is good.

To integrate the methane generation module with the biodegradation module, only three model parameters were available in the literature (Table S2); others had to be estimated from experimental data (Tables 2 and S4). Using these calculated values we applied the full stoichiometric model to methane measurements from a suite of experiments analogous to but independent of those used to estimate the parameters. Specifically, the $CH_4$ measurements were acquired during long-term incubation of MFT samples from Syncrude, CNUL and CNRL with their cognate diluents (Table S1, Siddique et al., 2015, Mohamad Shahimin and Siddique, 2017a, respectively). Figure 1 shows that the model predicted methane generation very well for all three types of MFT over long incubation times (> 4 yr incubation for CNUL and CNRL cultures). Additional modeling of Syncrude MFT with mixtures of *n*-alkane or monoaromatic components of its diluent (rather than whole diluent) also showed very good methane prediction (Fig. S6).

### 3.2 Model evaluation and comparison to previous models

Goodness-of-fit analysis of the stoichiometric model was calculated using NMSE (Table 3) and showed excellent fit, ranging from 0.81 – 0.98 for the three combinations of MFT and diluent. These NMSE results indicate that the integrated biodegradation and $CH_4$ methanogenesis modules faithfully capture the behaviour of independent laboratory cultures and that the stoichiometric model is sufficiently flexible to accommodate different inocula and substrates over long incubation periods.

The new stoichiometric model was then compared with the previous zero- and first-order kinetic models, as performed previously (Siddique et al., 2008), using the current data set. To this end, we first estimated the zero- and first-order kinetic model-related parameter values for the labile hydrocarbons that were not considered by Siddique et al. (2008) (Table S5). Figures 1 and S6, and Table 3 show that the stoichiometric model provides improved predictions over the previous models for describing $CH_4$ biogenesis from Syncrude MFT and whole naphtha or its components, and is far superior (matching closely with the measured methane) to the simpler models for the CNUL MFT–paraffinic diluent and for CNRL–naphtha combinations, neither of which were available for the previous modeling study. The improved fit regarding lag time and extent of $CH_4$ production, and the improved NMSE values suggest that the stoichiometric model, which is based on laboratory cultures, would be useful for modeling in situ $CH_4$ production from different OSTP and EPL.

### 3.3 Quantitative comparison of stoichiometric model predictions to measured cumulative $CH_4$ field emissions

To evaluate the feasibility of applying this model based on laboratory cultures to field emissions of $CH_4$, we compared the reported measured volumes of $CH_4$ emitted from the surfaces of OSTPs with cumulative $CH_4$ masses predicted by our model. Table 4 shows the comparison between the reported measured methane emissions from OSTPs in 2016 and 2017 and the maximum theoretical $CH_4$ yield predicted by our model based on the estimated diluent entering OSTPs (Table S6) for 2016 and 2017. The stoichiometric model predictions are 50-55 % of the measured emissions from Syncrude MLSB and 77-95% of the measured emissions from CNRL OSTP in both years. For CNUL where paraffinic solvent is used, the model predictions were 48% of the measured emissions in 2017 but only 17% of the emissions in 2016. This latter difference may be attributed to markedly greater methane emission data from CNUL OSTP reported in 2016 compared to all other OSTPs (Tables 4 and S5). The overall trend is very clear that the model predicted about 50% of emissions from Syncrude and CNUL OSTP and >75% of emissions from CNRL OSTP. This likely reflects the diluent compositions, with only ~40% of fugitive Syncrude and CNRL naphtha diluent being considered labile versus ~60% of CNUL paraffinic diluent, based on the mass of known biodegradable hydrocarbons in the diluents (Table S7).

This difference between predicted and measured $CH_4$ masses suggests that (other than possible inaccuracies associated with field measurements)  there are other endogenous carbon sources present in OSTP that support methanogenesis but are not currently accounted for by the model. Such possible sources include (but are not limited to): (1) additional labile diluent hydrocarbons not yet identified in our laboratory incubations and therefore not included in the model; (2) recalcitrant hydrocarbons deposited in previous years (and therefore not included in the annual $C_i^{in}$ model input) that are slowly degraded as the community adapts to residual naphtha after depletion of the labile hydrocarbons in lower strata, e.g., some *iso*-alkanes and cycloalkanes having extremely long lag times or slow degradation rates (e.g., Abu Laban et al., 2015); (3) slowly-degradable metabolites produced historically during incomplete biodegradation of hydrocarbon or from non-hydrocarbon carbon substrates; (4) organic matter associated with clays in oil sands ores  (Sparks et al., 2003); (5) minor labile components of bitumen e.g., high molecular weight *n*-alkanes (Oberding and Gieg, 2018); and (6) organic additives used in ore processing and deposited with tailings, e.g., citrate that is used as an amendment in some OSTPs (Foght et al., 2017) and is a potentially large source of unaccounted $CH_4$ in CNUL MRM. Another explanation for larger masses of measured emissions is the delayed, stochastic release of methane produced years ago from labile HCs that is 'trapped' in lower strata of MFT (Guo, 2009) until (1) suitably-sized and -oriented channels are created (e.g., by microbial activity, Siddique et al., 2014) and/or (2) cumulative gas voids reach critical buoyancy and rise from deep tailings, and/or (3) MFT strata are disturbed by some physical activity in the pond (e.g., moving deposition pipes, transferring MFT to new pits, etc.) allowing escape of gas.

There is good agreement between the model predictions and measured field emissions despite the obvious reasons of discrepancy discussed above. However, additional qualitative factors must be addressed to expand the developed model to in situ predictions while keeping in mind the inherent differences between laboratory cultures and field operations: (1) cultures are incubated with a single input of hydrocarbons, i.e., in "batch mode" with finite $C_i^{in}$, whereas the upper strata of OSTP receive ongoing input of diluent, i.e., "continuous mode" where $C_i^{in} > 0$. The laboratory cultures are more analogous to EPL, where $C_i^{in} = 0$ or to the lower strata of OSTP to which fresh diluent deposited at the surface cannot effectively diffuse and where, essentially, $C_i^{in} = 0$. (2) As discussed above, anaerobic biodegradation kinetics are currently available for only 18 hydrocarbons in cultures, whereas additional constituents of whole diluent and possibly a small subset of bitumen constituents may be susceptible to biodegradation in situ. Restriction of the parameter $C_i$ to the current 18 hydrocarbons would likely cause the model to under-estimate methane production in situ. Selective depletion of naphtha constituents with depth in OSTP has been observed qualitatively (Figure S2 in Foght et al., 2017) and such information could be used in future to expand the substrate range of the stoichiometric model and better represent in situ biodegradation. (3) The model currently includes a variable for lag time ($\lambda$), the time elapsed between addition of hydrocarbon and appearance of measureable $CH_4$. In fact, lag times of 5-15 years were observed between the inauguration of OSTP and the first observation of ebullition at the pond surface (Foght et al., 2017), likely reflecting the time required for establishment of efficient methanogenic communities. However, this variable is likely relevant only to laboratory studies, due to disruption of the microbial consortia during initiation of the cultures, and to newly established OSTP and EPL when transfer of tailings begins. After onset of $CH_4$ production, OSTP subsequently do not exhibit any apparent lag phases because of continuous diluent input and $\lambda=0$ in situ. (4) Small scale culture bottles facilitate release of $CH_4$ from MFT to the headspace for measurement compared with static deep strata in OSTP and EPL that experience physical retention of GHG as methane voids (Guo, 2009). That is, the model predicts $CH_4$ production based on 100% release from MFT; the proportion of gas released to the pond surface versus that retained under hydraulic pressure in situ is not a component of the model. (5) Methanogenesis depends completely upon the microbial community composition, which is complex (An et al., 2013) and specific to each OSTP and EPL (Wilson et al., 2016), and may diverge from cultured communities during incubation. Although some diversity data exist both for cultures and various MFT, the model does not include parameters to account for the presence or abundance of 'keystone' microbial species because, in tailings, such species currently are incompletely known or identified. Significant efforts in research and testing would be required to integrate microbial community analysis into any $CH_4$ model for oil sands operations. (6) Finally, the model does not currently include parameters that reflect potential changes to ore processing or OSTP practices such as subtle alterations in diluent composition, intermittent deposition of chemicals from related processes (e.g., ammonium; Foght et al., 2017), changes in froth treatment water temperature, etc.

## 3.4 Qualitative test of model prediction

Despite the inferred shortcomings of applying the model to field predictions, and in anticipation of acquiring in situ measurements to provide parameters for use in future for field modeling, it is possible to conduct a qualitative test of the stoichiometric model to determine whether it predicts expected trajectories under different expected field scenarios, e.g., limiting $C_T$ and/or $N_A$ conditions. Whereas cultures receive hydrocarbons in excess of instantaneous microbial demand at the beginning of incubation, as do the upper strata of active OSTP, labile carbon may become limiting in lower (older) strata of OSTP and eventually in EPL and cultures, where diluent is not replenished. Similarly, cultures initially receive a very small but finite amount of soluble nitrogen and have a headspace of $N_2$ gas (which may serve as a nitrogen source for tailings microbiota; Collins et al., 2016) but the lower strata of OSTP and EPL have no obvious input of biologically available nitrogen ($N_A$). Therefore this nutrient (or others, currently unidentified) may become limiting with time. Thus, challenging a model developed using culture data with scenarios reflecting in situ conditions should reveal the strength of the model. Phase plane analyses of eight forms of potential solutions of the stoichiometric model are shown in Figures S7 and S8 and described in Supplemental Material section S3. The model outputs describe the expected trajectories of OSTP and EPL under carbon and/or nitrogen limitation, solving for biomass and total carbon in the system with time, i.e., the sum of all microbial activity in situ. The predicted behaviour of OSTP with continuous diluent input differs from EPL with no additional hydrocarbon input, and the effect of limiting nutrient (nitrogen) also changes the ultimate endpoints of biomass and carbon in the two scenarios. These outputs qualitatively support the validity of the model as well as indicating that the stoichiometric model could be used to predict specific OSTP and EPL behaviour, to predict the volumes of 'legacy' $CH_4$ from OSTP and long-term duration of $CH_4$ production in situ (particularly from EPL), and to influence decisions about oil sands reclamation strategies. If additional in situ model parameters are acquired, the model can be further refined to improve predictive power.

## 4. Conclusions

The stoichiometric model represents a significant advance over previous zero- and first-order kinetic models, particularly because it predicts well the behaviour of tailings from different operators using distinct diluents that may support different rates of $CH_4$ production or may ultimately generate greater $CH_4$ emissions. Application of the model to in situ $CH_4$ production is still hampered by limited experimental data and field measurements; some of these gaps may be alleviated as relevant in situ data are acquired and when future anaerobic studies provide both evidence for susceptibility of additional hydrocarbons to biodegradation and more precise values for model parameters. The model is sufficiently flexible that additional parameters can be added to the modules as laboratory or field data become available. Until such time, the stoichiometric model should assist regulators and oil sands operators in qualitatively assessing long-term GHG emissions from oil sands tailings deposits and EPL reclamation sites.

## Appendix A. Supplementary Material

This manuscript is accompanied by Supplementary Material comprising stability analysis of our System, eight tables (Tables S1-S8) and eight figures (Figure S1-S8).

## ACKNOWLEDGMENTS

We acknowledge support from NSERC Discovery Grants (TS, JF, HW and MAL), NSERC Postdoctoral Fellowship (#PDF-502490-2017; JK) and a Canada Research Chair (MAL). In addition, JK thanks DIMACS for providing space to conduct the analyses (partially enabled through support from the National Science Foundation under grant #CCF-1445755.)

**Table 1:** Definition of terms used in model development

| Term | Definition |
|---|---|
| $C_i$ | mass of individual labile hydrocarbons in the system, where $i$=1…$n$, assuming $n$ labile hydrocarbons in system * |
| $C_i^{in}$ | mass of $C_i$ inflow to the system |
| $C_T$ | total mass of labile (biodegradable) hydrocarbon in the system (i.e., the sum of all $C_i$) |
| $\mu$ | specific microbial growth rate of microbes (Bacteria and Archaea) supported by $C_T$ |
| $\mu_i$ | specific microbial growth rate supported by each labile hydrocarbon $C_i$ |
| $N_T$ | total mass of nitrogen in the system |
| $N_A$ | mass of $N_T$ that is biologically available § |
| B | total biomass of living microbes |
| b | biomass of dead microbes |
| $\beta_i$ | the proportion of $C_i$ contained in dead biomass that is available for microbial recycling |
| $\theta$ | the ratio of nitrogen to carbon associated with microbial biomass B |
| $r$ | proportionality constant defining efficiency of conversion of $C_T$ to B |
| $r_i$ | proportionality constant defining efficiency of conversion of each $C_i$ to B; $r_i$ = B / $C_i$ consumed |
| $\lambda_i$ | lag period before the onset of biodegradation of each $C_i$ |
| $d$ | microbial cell death rate |
| $K_f$ | $N_A$-dependent half-saturation constant |
| $K_{gi}$ | $C_i$-dependent half-saturation constant |
| $\Gamma_i$ | expected yield of $CH_4$ from biodegradation of one mole of $C_i$ |
| $G_i$ | Total $CH_4$ and $CO_2$ generated from the biodegradation of $C_i$ |
| $\eta$ | fraction of sum of $\Gamma_i$ for all $i$, yielded by biodegradation of $C_T$; i.e., methane bioconversion efficiency factor |
| $\eta_i$ | fraction of $\Gamma_i$ yielded by biodegradation of each $C_i$ |

\*, in developing the current model, we considered 18 specific hydrocarbons present in naphtha and paraffinic diluents (see Table 2)

§, e.g., nitrate, nitrite, ammonium, dinitrogen ($N_2$ gas), labile organic N compounds (e.g., macromolecules in biomass), but not complex molecules (e.g., resins found in bitumen)

**Table 2**: List of 18 labile diluent hydrocarbons used in model development, sources of data and type of tailings used to generate data for the biodegradation module and to estimate model parameter values, and the model parameters estimated using those data (see Table S4 for parameter definitions and values).

| Hydrocarbon | Source of data | Type of tailings | Parameters estimated from the data |
|---|---|---|---|
| *n*-Alkanes | | | |
| $C_5$ | Mohamad Shahimin et al. (2016) | CNUL | $K_{g_{C_5}}$ and $C_5$-lag |
| $C_6, C_7, C_8, C_{10}$ | Siddique et al. (2006) | Syncrude | B(0), $K_f$, $N_T$, $K_{g_{C_6}}$, $K_{g_{C_7}}$, $K_{g_{C_8}}$, $K_{g_{C_{10}}}$, $C_6$-lag, $C_7$-lag, $C_8$-lag and $C_{10}$-lag. |
| $C_9$ | Table S1 | Syncrude | $K_{g_{C_9}}$ and $C_9$-lag |
| *iso*-Alkanes * | | | |
| $2\text{-MC}_6^{\S}$, $3\text{-MC}_6$, $2\text{-MC}_7$, $4\text{-MC}_7$, $2\text{-MC}_8$, $3\text{-MC}_8^{\S}$, $2\text{-MC}_9^{\S}$ | Siddique et al., unpublished | Syncrude | $K_{g_{3-MC_6}}$, $K_{g_{2-MC_7}}$, $K_{g_{4-MC_7}}$, $K_{g_{2-MC_8}}$, $3\text{-MC}_6$–lag, $2\text{-MC}_7$-lag, $4\text{-MC}_7$-lag, and $2\text{-MC}_8$-lag |
| $2\text{-MC}_5$ | Mohamad Shahimin and Siddique (2017a) | CNUL | $K_{g_{2-MC_5}}$ and $2\text{-MC}_5$-lag |
| Monoaromatics | | | |
| Toluene, *o*-Xylene, *m*- plus *p*-Xylene | Siddique et al. (2007) | Syncrude | $K_{g_{toluene}}$, $K_{g_{o-xylene}}$, $K_{g_{mp-xylene}}$, toluene-lag, *o*-xylene-lag, and *m,p*-xylene-lag |

\* M denotes a methyl group; i.e., $2\text{-MC}_6$ is 2-methylhexane, etc. See Methods section 2.3.1 for full list of abbreviations

§ The values of model parameters $K_g$ and lag for $2\text{-MC}_6$, $3\text{-MC}_8$ and $2\text{-MC}_9$ are not available from empirical studies and are assumed to be the same as those for $3\text{-MC}_6$, $2\text{-MC}_8$ and $2\text{-MC}_8$, respectively, due to their similar molecular weights.

**Table 3**: Normalized mean square error (NMSE) analysis of model predictions and measured $CH_4$ production from laboratory cultures comprising three MFT samples incubated with their cognate diluents. The zero- and first-order models were implemented as described by Siddique et al. (2008) using data reported in the current study. See Figures 1 and S6 for graphical comparison of model outputs.

| | NMSE values | | |
|---|---|---|---|
| | MFT source and diluent type | | |
| | Syncrude | CNUL | CNRL |
| Model | Naphtha diluent | Paraffinic diluent | Naphtha diluent |
| Zero-order | -0.28 | -1.00 | -1.10 |
| First-order | -0.65 | 0.82 | 0.61 |
| Stoichiometric | 0.81 | 0.98 | 0.97 |

**Table 4:** Comparison of cumulative field measurements of $CH_4$ emissions in 2016 and 2017 in three OSTP versus stochiometric model predictions of cumulative in situ $CH_4$ emissions from those OSTP.

| Operator and OSTP (date) | Field measurements of $CH_4$ emissions (moles x $10^6$) * | Stochiometric model predictions of methane emissions (moles x $10^6$) | Proportion of field emissions predicted by model (%) § |
|---|---|---|---|
| Syncrude MLSB (2016) | 1191 | 656 | 55 |
| Syncrude MLSB (2017) | 991 | 492 | 50 |
| CNRL Horizon (2016) | 336 | 321 | 95 |
| CNRL Horizon (2017) | 599 | 459 | 77 |
| CNUL MRM (2016) | 2634 | 445 | 17 |
| CNUL MRM (2017) | 1051 | 506 | 48 |

* Unpublished surface flux measurements (Government of Alberta; raw data available upon request), reported as litres and converted to moles at standard temperature and pressure

§ for detailed calculations see Table S8

**FIGURE LEGEND**

**Figure 1**: Comparison of $CH_4$ production predicted by the stoichiometric model versus $CH_4$ measured in laboratory cultures independent of those used to generate the stoichiometric model and parameters (Table S4). Methane measurements (diamond symbols) are from cultures comprising: (A), Syncrude MFT incubated with its naphtha diluent (B), CNUL MFT incubated with its paraffinic diluent; and (C), CNRL MFT incubated with its naphtha diluent. Solid lines represent the stoichiometric model prediction; dashed lines and dotted lines respectively represent predictions made by applying the previous zero-order and first-order models ( Siddique et al., 2008) to the independent data set.   The parameters values used in simulating the zero-order and first-order models were obtained from Siddique et al. (2008) and Table S5.

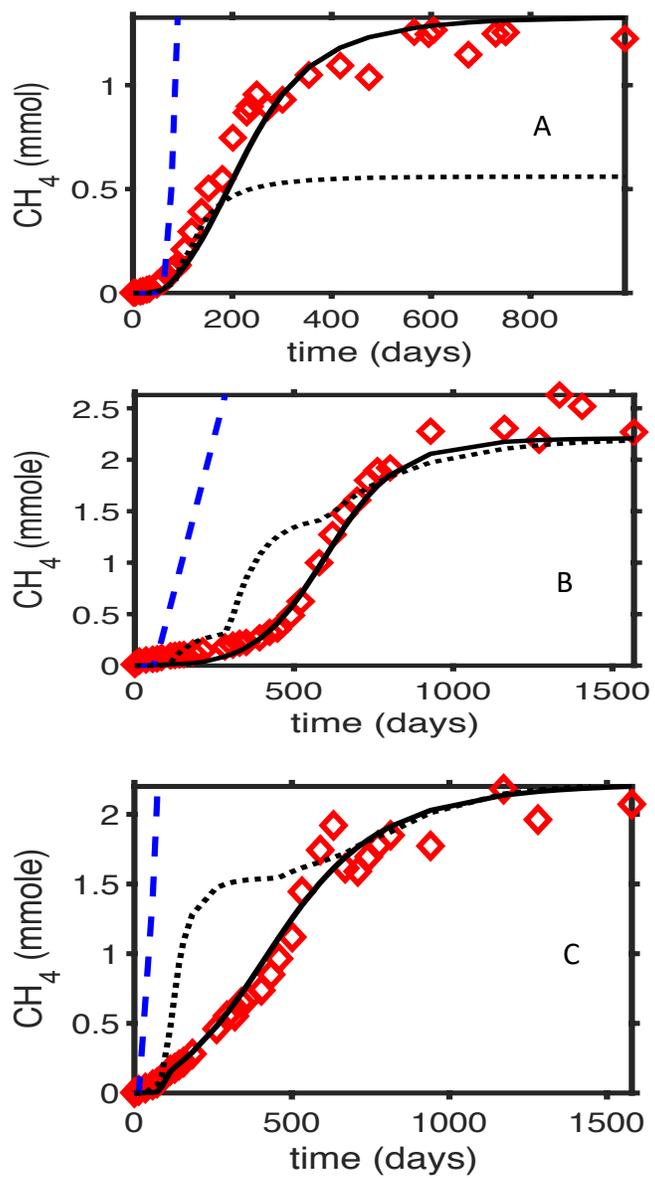

**Figure 1**

# Appendix A:

# Second-generation stoichiometric mathematical model to predict methane emissions from oil sands tailings

5    Jude Kong[1,2], Hao Wang[2*†], Tariq Siddique[3*‡], Julia Foght[4], Kathleen Semple[4], Zvonko Burkus[5],

and Mark Lewis[2,4]

[1]Center for Discrete Mathematics and Theoretical computer Science, 96 Frelinghuysen Road Piscataway, NJ 08854-8018, USA
10   [2]Department of Mathematical and Statistical Sciences, University of Alberta, Edmonton, AB T6G 2G1, Canada
[3]Department of Renewable Resources, University of Alberta, Edmonton, AB T6G 2G7, Canada
[4]Department of Biological Sciences, University of Alberta, Edmonton, AB T6G 2E9, Canada
[5]Alberta Environment and Parks, Government of Alberta, Edmonton, Canada

Corresponding authors' emails:

[*†] Mathematical approach (Hao Wang); hao8@ualberta.ca

[*‡] Biological approach (Tariq Siddique); tariq.siddique@ualberta.ca

20   The following Supplementary Material contains the mathematical analysis of the system of equations (2), eight tables (Tables S1- S8) and eight figures (Figures S1-S8).

## S1. Brief description of MFT laboratory culture methods used to generate

25   data for model development and testing

Details of laboratory culture preparation can be found in published papers ( Mohamad Shahimin et al., 2016; Mohamad Shahimin and Siddique, 2017a, 2017b, Siddique et al., 2007, 2006). Briefly and very generally, bulk samples of MFT were dispensed anaerobically into small serum bottles (microcosms) in replicate (typically triplicates) amended with an equal volume of sterile

30  methanogenic medium comprising inorganic salts, trace vitamins, a redox indicator and sulfide as a reducing agent, but lacking organic carbon, and sealed under an atmosphere of 80% $O_2$-free $N_2$, balance $CO_2$. The microcosms were allowed to incubate stationary in the dark at room temperature (ca. 22°C) for 2 weeks to acclimate, then the headspace was flushed with $O_2$-free $N_2$ plus $CO_2$ to remove any $CH_4$ produced from endogenous substrates. The microcosms were then

35  amended by injecting neat diluent supplied by the operator, or in one case defined mixtures of pure hydrocarbon constituents of the diluent (i.e., mixtures of *n*-alkanes or monoaromatics; Figure S6). During incubation headspace gases were sub-sampled at intervals for analysis by gas chromatography to determine cumulative $CH_4$ production. Likewise the MFT slurry was sub-sampled at intervals to analyze residual hydrocarbons using gas chromatography with mass

40  spectrometry and thereby to calculate biodegradation by difference. Control microcosms containing MFT that had been heat-sterilized using an autoclave were included with each experiment to account for any abiotic losses of hydrocarbons.

## S2. Model development details

45  ### *S2.1 Mathematical analysis of the biodegradation module*

Here, a basic mathematical analysis of the system of equations (2) is provided. First we let $C_T$ to represent the sum of all the labile hydrocarbons in the system and the sum of all $C_i^{in}$ to be $C_T^{in}$.. We assume that $\lambda_i = 0$, for all i=1,2,3..n. This leads to a system of two differential equations.

To simplify our phase plane analysis in a meaningful way, we adjusted the second differential by introducing a new variable:

$A = \frac{B}{r} + C_T$. 'A' represents the sum of the total carbon available in the system and bacterial biomass. We assume that $f, g$ are linear and find their linear approximations:

$$f(N_T - \theta B) \approx f(0) + f'(0)(N_T - \theta B)$$

$$\Rightarrow f(N_T - \theta B) \approx \frac{N_T - \theta B}{K_f}$$

$$g\left(A - \frac{B}{r}\right) \approx g(0) + g'(0)\left(A - \frac{B}{r}\right)$$

$$\Rightarrow g\left(A - \frac{B}{r}\right) \approx \frac{A - \frac{B}{r}}{K_g}$$

We thus have the following system in which only one of the two differential equations has a minimum function, greatly simplifying the analysis:

$$\dot{A} = \frac{r-1}{r} dB + C_T^{in} = F(B) \qquad\qquad (S1)$$

$$\dot{B} = \mu B \min\left\{f(N_T - \theta B), g\left(A - \frac{B}{r}\right)\right\} - dB = BG(A, B).$$

Next, we look at the stability analysis of the system. For this purpose, we construct a phase plane of the system, (i.e. a graph of the solution trajectories mapped out by points *(A(t),B(t))* as t varies over $(\infty, +\infty)$) in order to identify the steady state solutions. We call $F(B) = 0$ and $G(A, B) = 0$ (the lines on which trajectories are horizontal or vertical) the nullclines of system of equations (S1). The steady state solutions are the points where the nullclines (but not different branches of the same nullcline) cross each other. For the stability of the steady states, we compute the Jacobian matrix corresponding to each equilibrium point $J(A^*, B^*)$, where $(A^*, B^*)$ is a given equilibrium point. We use the sign of the trace and determinant of $J(A^*, B^*)$ to determine the

70   nature of the given equilibrium point. Let D = det $J(A^*, B^*)$ and $T_r$ = trace $J(A^*, B^*)$. Note that:

   1) If $D < 0$, the eigenvalues of J(A*,B*) are real and of opposite signs, and the phase

   portrait is a saddle (which is always unstable).

   2) If $0 < D < \frac{T_r^2}{4}$, the eigenvalues of $J(A^*, B^*)$ are real, distinct, and of the same sign, and

   the phase portrait is a node, stable if $T_r < 0$ and unstable if $T_r > 0$.

75   3) If $0 < T_r^2 < D$, the eigenvalues of $J(A^*, B^*)$ are neither real nor purely imaginary, and

   the phase portrait is a spiral, stable if $T_r < 0$ and unstable if $T_r > 0$. Using this idea, we

   carried out the analysis as follows:

### S2.2 Stability Analysis of OSTP system ($C_T^{in} \neq 0$)

80   **Steady states:**

**A-Nullclines:**

$\dot{A} = 0, \implies B = \frac{rC_T^{in}}{d(1-r)}.$

**B-Nullclines:**

$$\dot{B} = 0, \implies B = 0 \ or \ G(A, B) = 0.$$

85   $$G(A, B) = 0, \implies \begin{cases} B = Ar - \frac{dk_g r}{\mu} \ if \ \frac{N_T - \theta B}{k_f} > \frac{A - \frac{B}{r}}{k_g} \\ B = \left(N_T - \frac{dk_f}{\mu}\right)\frac{1}{\theta} \ if \ \frac{N_T - \theta B}{k_f} < \frac{A - \frac{B}{r}}{k_g} \end{cases}$$

**Case 1:** Suppose $\theta - \frac{k_f}{k_g r} > 0$, then

$$G(A,B) = 0, \Longrightarrow \begin{cases} B = Ar - \dfrac{dk_g r}{\mu} \; if \; B < \left(N_T - \dfrac{Ak_f}{k_g}\right)\left(\dfrac{k_g r}{\theta k_g r - k_f}\right) \\ B = \left(N_T - \dfrac{dk_f}{\mu}\right)\dfrac{1}{\theta} \; if \; B > \left(N_T - \dfrac{Ak_f}{k_g}\right)\left(\dfrac{k_g r}{\theta k_g r - k_f}\right) \end{cases}$$

**Case 1.1:** If $C_T^{in} > \dfrac{d(1-r)}{r\theta}\left(N_T - \dfrac{dk_f}{\mu}\right)$, there will be no intersection between the $A$ and $B$- nullclines as shown in Panel A of Figure S7. Hence the system will have no equilibrium point.

**Case 1.2:** If $C_T^{in} < \dfrac{d(1-r)}{r\theta}\left(N_T - \dfrac{dk_f}{\mu}\right)$, the two nullclines will intersect at one unique point $E_1 =$

$\left(\dfrac{\mu C^{in} + d^2 k_g(1-r)}{d(1-r)\mu}, \dfrac{rC_T^{in}}{d(1-r)}\right)$ as shown in Panel B of Figure S7. Hence if

$C_T^{in} < \dfrac{d(1-r)}{r\theta}\left(N_T - \dfrac{dk_f}{\mu}\right)$, the system will have a unique internal equilibrium point $E_1$.

**Case 1.3:** If $C_T^{in} = \dfrac{d(1-r)}{r\theta}\left(N_T - \dfrac{dk_f}{\mu}\right)$, the two nullclines will intersect on the line $\left\{\left(A, \left(T - \dfrac{dk_f}{\mu}\right)\dfrac{1}{\theta}\right) : A > \dfrac{1}{r}\left[\left(T - \dfrac{dk_f}{\mu}\right)\dfrac{1}{\theta} + \dfrac{dk_g r}{\mu}\right]\right\}$ as can be seen in Panel A of Figure S8. Consequently, If

$C_T^{in} = \dfrac{d(1-r)}{r\theta}\left(N_T - \dfrac{dk_f}{\mu}\right)$, the system will have an infinite number of equilibrium points $E_2 =$

$\left\{\left(A, \left(N_T - \dfrac{dk_f}{\mu}\right)\dfrac{1}{\theta}\right) : A > \dfrac{1}{r}\left[\left(N_T - \dfrac{dk_f}{\mu}\right)\dfrac{1}{\theta} + \dfrac{dk_g r}{\mu}\right]\right\}$

**Case 2:** Suppose $\theta - \dfrac{k_f}{k_g r} < 0$, then

$$G(A,B) = 0, \Longrightarrow \begin{cases} B = Ar - \dfrac{dk_g r}{\mu} \; if \; B > \left(N_T - \dfrac{Ak_f}{k_g}\right)\left(\dfrac{k_g r}{\theta k_g r - k_f}\right) \\ B = \left(N_T - \dfrac{dk_f}{\mu}\right)\dfrac{1}{\theta} \; if \; B < \left(N_T - \dfrac{Ak_f}{k_g}\right)\left(\dfrac{k_g r}{\theta k_g r - k_f}\right). \end{cases}$$

Note that the slope of the line $B = Ar - \frac{dk_g r}{\mu}$ is less than that of $B = \left(N_T - \frac{Ak_f}{k_g}\right)\left(\frac{k_g r}{\theta k_g r - k_f}\right)$, since $\frac{k_f}{k_f - \theta k_g r} > 1$. Therefore, the point where the line $B = Ar - \frac{dk_g r}{\mu}$ intersects the $A$-axis, $\frac{dk_g}{\mu}$, must be less than $\frac{Tk_g}{k_f}$, the point where the $B = \left(N_T - \frac{Ak_f}{k_g}\right)\left(\frac{k_g r}{\theta k_g r - k_f}\right)$ intersect the A-axis, for the two lines to intersect on the first quadrant.

**Case 2.1:** If $C_T^{in} > \frac{d(1-r)}{r\theta}\left(N_T - \frac{dk_f}{\mu}\right)$, as with Case 1.1, there will be no intersection between the $A$ and $B$-nullclines as shown in Panel B of Figure S8. Hence the system will have no equilibrium point.

**Case 2.2:** $C_T^{in} < \frac{d(1-r)}{r\theta}\left(N_T - \frac{dk_f}{\mu}\right)$, the two nullclines will intersect at one unique point $E_3 =$

$\left(\frac{\mu C^{in} + d^2 k_g(1-r)}{d(1-r)\mu}, \frac{rC_T^{in}}{d(1-r)}\right)$ as shown in Panel C of Figure S8. Hence if $\theta\, C_T^{in} < \frac{d(1-r)}{r\theta}\left(N_T - \frac{dk_f}{\mu}\right)$, the system will have a unique internal equilibrium point $E_3$.

**Case 2.3:** If $C_T^{in} = \frac{d(1-r)}{r\theta}\left(N_T - \frac{dk_f}{\mu}\right)$, the two nullclines will intersect on the line $\Big\{\Big(A, \Big(N_T -$

$\frac{dk_f}{\mu}\Big)\frac{1}{\theta}\Big) : A > \frac{1}{r}\Big[\Big(N_T - \frac{dk_f}{\mu}\Big)\frac{1}{\theta} + \frac{dk_g r}{\mu}\Big]\Big\}$ as shown in Panel D of Figure S8. Thus, If $C_T^{in} =$

$\frac{d(1-r)}{r\theta}\Big(N_T - \frac{dk_f}{\mu}\Big)$, the system will have an infinite number of equilibrium points $E_4 =$

$\Big\{\Big(A, \Big(N_T - \frac{dk_f}{\mu}\Big)\frac{1}{\theta}\Big) : A > \frac{1}{r}\Big[\Big(N_T - \frac{dk_f}{\mu}\Big)\frac{1}{\theta} + \frac{dk_g r}{\mu}\Big]\Big\}$

Thus an OSTP system may have 0, 1, or an infinite number of equilibrium points depending on the volume of fresh labile hydrocarbons input into the system, $C_T^{in}$. If $C_T^{in} > \frac{d(1-r)}{r\theta}\left(N_T - \frac{dk_f}{\mu}\right)$, the system will have no equilibrium point; if $C_T^{in} < \frac{d(1-r)}{r\theta}\left(N_T - \frac{dk_f}{\mu}\right)$, it will have one

125  unique equilibrium point , $\left(\frac{\mu C_T^{in} + d^2 k_g(1-r)}{d(1-r)\mu}, \frac{rC_T^{in}}{d(1-r)}\right)$; and if $\frac{rC_T^{in}}{d(1-r)} = \left(N_T - \frac{dk_f}{\mu}\right)\frac{1}{\theta}$, it will have an infinite number of equilibrium points given by $\left\{\left(A, \left(N_T - \frac{dk_f}{\mu}\right)\frac{1}{\theta}\right): A > \frac{1}{r}\left[\left(N_T - \frac{dk_f}{\mu}\right)\frac{1}{\theta} + \frac{dk_g r}{\mu}\right]\right\}$.

### S2.2.1 Stability of equilibrium points in OSTP scenario:

130  To determine the local stability of the equilibria above, we consider the Jacobian matrix of System of equations (S1),

$$J(A,B) = \begin{pmatrix} 0 & \frac{(r-1)d}{r} \\ BG_A(A,B) & G(A,B) + BG_B(A,B) \end{pmatrix} \quad (S1.)$$

Where

$$G(A,B) = \begin{cases} \dfrac{\mu\left(A - \frac{B}{r}\right)}{k_g} - d \ if \ \dfrac{N_T - \theta B}{k_f} > \dfrac{A - \frac{B}{r}}{k_g} \\ \dfrac{\mu(N_T - \theta B)}{k_f} - d \ if \ \dfrac{N_T - \theta B}{k_f} < \dfrac{A - \frac{B}{r}}{k_g}, \end{cases}$$

135

$$G_A(A,B) = \begin{cases} \dfrac{\mu}{k_g} \ if \ \dfrac{N_T - \theta B}{k_f} > \dfrac{A - \frac{B}{r}}{k_g} \\ 0 \ if \ \dfrac{N_T - \theta B}{k_f} < \dfrac{A - \frac{B}{r}}{k_g} \end{cases}$$

and

$$G_B(A, B) = \begin{cases} \dfrac{-\mu}{k_g} \; if \; \dfrac{N_T - \theta B}{k_f} > \dfrac{A - \dfrac{B}{r}}{k_g} \\[4mm] \dfrac{-\theta}{k_f} \; if \; \dfrac{N_T - \theta B}{k_f} < \dfrac{A - \dfrac{B}{r}}{k_g} \end{cases}$$

**Stability of $E_1$:**

$$J(E_1) = \begin{pmatrix} 0 & \dfrac{(r-1)d}{r} \\[4mm] \dfrac{\mu r C_T^{in}}{k_g d(1-r)} & \dfrac{-C_T^{in}\mu}{d(1-r)k_g} \end{pmatrix} \quad (S2.)$$

Since $det\big(J(E_1)\big) = \frac{\mu c_T^{in}}{k_g}$ is greater than zero and $T_r\big(J(E_1)\big) = \frac{-c_T^{in}\mu}{d(1-r)k_g} < 0$, this implies that both eigenvalues of $J(E_1)$ have negative real parts. Hence $E_1$ is a locally stable equilibrium point. It is easy to see that $E_1$ is a stable spiral.

**Stability of $E_2$:**

$$J(E_2) = \begin{pmatrix} 0 & \dfrac{(r-1)d}{r} \\[4mm] 0 & \dfrac{-r C_T^{in}\mu\theta}{d(1-r)k_f} \end{pmatrix} \quad (S3.)$$

$det\big(J(E_2)\big) = 0$ and $T_r\big(J(E_2)\big) = \frac{-r c_T^{in}\mu\theta}{d(1-r)k_f} < 0$.

Since the $T_r(J(E_2))$ is negative and $det(J(E_2))$ is zero, one eigenvalue is zero and the other is negative. Thus $E_2$ is a line of locally asymptotically stable equilibrium points. Hence both the internal equilibrium point $E_1$ and the line of equilibrium points $E_2$ are locally asymptotically stable.

***S2.2.2 End pit lake scenario ($C_T^{in} = 0$):***

155    **Steady states:**

***A*-Nullclines:**

$$\dot{A} = 0 \implies B = 0$$

***B*-Nullclines:**

$$\dot{B} = 0 \implies B = 0 \ or \ G(A, B) = 0.$$

160

Panels C and D of Figure S7 show that, irrespective of the slope of the line $B = Ar - \frac{dk_g r}{\mu}$, the $A$-and $B$-nullclines have an infinite number of intersections, given by $E_5 = \{(A, 0) : A \geq 0\}$. Thus for $C_T^{in} = 0$, system of equations (S1) has an infinite number of equilibrium points given by $E_5$.

165

**Stability of $E_5$:**

$$J(E_5) = \begin{pmatrix} 0 & \dfrac{(r-1)d}{r} \\ 0 & \dfrac{\mu A}{k_g} - d \end{pmatrix} \ (4.)$$

$det\big(J(E_5)\big) = 0$ and $T_r\big(J(E_5)\big) = \frac{\mu A}{k_g} - d$. If $A < \frac{dk_g}{\mu}$, $T_r(J(E_5))$ will be less than zero and hence $E_5$ will be asymptotically stable. On the other hand, if $A \geq \frac{dk_g}{\mu}$, then $T_r(J(E_5))$ will be greater

170    than 0 and thus $E_5$ will be a line of unstable equilibrium points.

## S3. Qualitative challenge of model prediction

Figures S7 and S8 show eight theoretical in situ scenarios presented as phase plane diagrams showing solutions for microbial biomass versus total carbon content (both unitless) under conditions of carbon or nitrogen limitation. The directional arrows account for time, nullclines define the vector fields, and nullcline intersections (fixed points) indicate regions where trajectories are horizontal or vertical; i.e., steady states. Panels S7A , S7B and S8A-S8D are relevant to the upper strata of OSTP where the input of labile hydrocarbon is continuous (i.e., $C_T^{in} > 0$) whereas Panels S7C and S7D represent an established EPL where labile carbon (as partially biodegraded diluent) enters the system with deposited MFT but is not replenished (i.e., $C_T^{in} = 0$) Furthermore, the availability of nitrogen ($N_A$) differs for each panel, as described below.

Let $C_0^{in}$, $C_1^{in}$ and $C_2^{in}$ denote sums of labile hydrocarbons with values $\left(N_T - \frac{dk_f}{\mu}\right)\frac{d(1-r)}{\theta r}$, $\left(\frac{C_T^{in}}{d(1-r)} + \frac{dK_g}{\mu}\right)$ and $\frac{1}{r}\left[\left(T - \frac{dk_f}{\mu}\right)\frac{1}{\theta} + \frac{dk_g r}{\mu}\right]$ respectively. Also, let $B_0$ and $B_1$ denote two different values of bacterial biomass. $B_0 = \left(N_T - \frac{dk_f}{\mu}\right)\frac{1}{\theta}$ and $B_1 = \frac{d(1-r)}{\theta r}$. Figures S7A and S8B show the predicted behaviour of OSTP in which the rate of input of hydrocarbons into the OSTP per unit time is $> C_0^{in}$. In this scenario, biomass moves towards $B_0$ (i.e., steady state). As biomass stabilizes, nitrogen becomes the limiting factor in microbial growth and thus Bacteria consume only the amount of hydrocarbon permitted by $N_A$. This leads to a accumulation of hydrocarbon in the system due to the continuous influx of diluent and inability of Bacteria to degrade all the carbon input. Such a scenario would require addition of $N_A$ to the ponds to achieve additional diluent consumption, if that was the management goal. Conversely, restricting $N_A$ in the pond should decrease $CH_4$ and $CO_2$ emissions although the potential for gas biogenesis would persist for an indefinite period. Figures S7B and S8D illustrates the case of an OSTP

195 where the rate of input of hydrocarbons into the OSTP per unit time is $< C_0^{in}$. In this case, biomass moves to a value of $B_1$ and total $C_T^{in}$ moves to $C_1^{in}$. Because the total labile hydrocarbon deposited into the pond per unit time $C_T^{in}$ is $< C_0^{in}$, carbon becomes the limiting factor for bacterial growth. Thus, biomass will increase to achieve a steady state at which carbon intake is maximized and all $C_T$ is degraded as it enters the system. This scenario requires a continuous

200 (but currently undiscovered) source of $N_A$ in the tailings or the addition of exogenous $N_A$, i.e., as a management practice. The final possible scenario in OSTP is that depicted in Figures S8A and S8C. As with the other two cases above, we are equally looking at the OSTP as defined by the continuous input of carbon. Here the rate of input of hydrocarbons into the OSTP per unit time is $C_0^{in}$. At this influx value per unit time, nitrogen would be the limiting element for microbial

205 growth. In this scenario, we have microbes growing to $B_0$, a point where they can maximize they nitrogen intake. Carbon in turn changes to a value that is greater than $C_2^{in}$.

The scenarios in Figures S7C and S7D simulate EPL conditions because $C_T^{in} = 0$. With extended time, $C_T$ will approach a minimum (theoretically zero) as $C_T$ is converted to $CH_4$ and

210 dead biomass is likewise degraded after labile hydrocarbons are depleted. Figure S7C describes a scenario where the ratio of the nitrogen carrying capacity to carbon carrying capacity of the pond is $< \theta r$. Since there is no supply of exogenous carbon to the system, when the Bacteria degrade all residual diluent, they ultimately have no carbon source other than dead biomass, which is converted to $CH_4$ and $CO_2$; eventually gas generation ceases in this closed system.

215 Figure S7D predicts the situation where the ratio of the nitrogen carrying capacity to carbon carrying capacity of the pond is $> \theta r$ but $C_T$ still approaches zero because of the complete conversion of $C_T$ and $\beta_T\, dB$ to gases, where $\beta_T$ is the proportion of $C_T$ contained in dead biomass

that is available for microbial recycling. Note that in the interim, biomass was greater than in

Figure S7C because of the continuous presence of $N_A$.

220

265

270

275 **Table:** Biodegradation and cumulative CH$_4$ production in cultures of Syncrude MFT incubated with Syncrude naphtha diluent.

| Hydrocarbon (mg L$^{-1}$) | Incubation period (days) | | | | | | | | | |
|---|---|---|---|---|---|---|---|---|---|---|
| | 28 | 77 | 142 | 216 | 249 | 271 | 365 | 475 | 605 | 730 |
| Toluene | 46.0 | 38.2 | 0 | 0 | 0 | 0 | 0 | 0 | 0 | 0 |
| Ethylbenzene | 19.0 | 21.6 | 15.2 | 0 | 0 | 0 | 0 | 0 | 0 | 0 |
| *m-,p*-Xylenes | 35.0 | 46.2 | 35.0 | 36.9 | 28.7 | 10.1 | 7.7 | 0 | 0 | 0 |
| *o*-Xylene | 14.0 | 17.7 | 11.3 | 0 | 0 | 0 | 0 | 0 | 0 | 0 |
| *n*-Hexane | 5.0 | 2.5 | 2.7 | 1.2 | 0.7 | 0.4 | 0.4 | 0.3 | 0.3 | 0 |
| *n*-Heptane | 34.0 | 18.2 | 13.9 | 6.5 | 3.7 | 2.3 | 1.0 | 2.6 | 0 | 0 |
| *n*-Octane | 46.0 | 30.2 | 23.9 | 13.8 | 8.0 | 4.5 | 2.5 | 0 | 0 | 0 |
| *n*-nonane | 15.0 | 15.2 | 6.2 | 3.5 | 1.3 | 0 | 0 | 0 | 0 | 0 |
| 2-Methylhexane (2-MC$_6$) | 10.0 | 6.8 | 6.0 | 6.9 | 6.4 | 5.3 | 4.7 | 5.7 | 2.7 | 1.6 |
| 3-Methylhexane (3-MC$_6$) | 12.0 | 8.2 | 7.7 | 6.7 | 5.5 | 2.4 | 3.2 | 2.7 | 1.9 | 1.9 |
| 2-Methylheptane (2MC$_7$) | 37.0 | 25.0 | 22.1 | 25.5 | 23.8 | 19.7 | 17.3 | 21.3 | 10.2 | 5.8 |
| 4-Methylheptane (4-MC$_7$) | 14.0 | 9.6 | 8.4 | 8.4 | 4.4 | 3.5 | 4.4 | 4.5 | 3.4 | 0 |
| **Cumulative CH$_4$ production (µmol) *** | 16 | 114 | 416 | 774 | 955 | 893 | 1049 | 1039 | 1266 | 1248 |

* Cumulative methane is calculated by subtracting CH$_4$ produced by parallel endogenous control cultures (i.e., MFT not receiving additional naphtha) from CH$_4$ measured in test cultures (MFT receiving naphtha).

280

**Table S2**: Literature values for selected microbial parameters in system of equations (2)

| Parameter * | Value Range | Unit | References |
|:---:|:---:|:---:|:---|
| $\mu$ | 1-4 | d$^{-1}$ | (Codeco and Grover, 2001; Connolly et al., 1992) |
| $r$ | 0.31-0.75 | – § | (Del Giorgio and Cole, 1998; Wang et al., 2009) |
| $\theta$ | $\dfrac{1}{9} - \dfrac{1}{4}$ | – § | (Sterner and Elser, 2002) |

\* see Table 1, main text, for parameter definitions
–, unitless parameters

285 **Table S3**: Normalized mean square error (NMSE) values obtained by comparing the simulated biodegradation kinetics (generated using the system of equations (2) and parameter values in Table S4) to published experimental data for the 15 labile hydrocarbons (Table 2).

| Hydrocarbon * | NMSE |
|---|---|
| *n*-Pentane | 0.92 |
| *n*-Hexane | 0.99 |
| *n*-Heptane | 0.99 |
| *n*-Octane | 0.99 |
| *n*-Nonane | 0.98 |
| *n*-Decane | 0.99 |
| Toluene | 1.00 |
| *o*-Xylene | 1.00 |
| *m*- plus *p*-Xylene | 0.99 |
| 2-Methylpentane | 1.00 |
| 3-Methylhexane | 0.99 |
| 2-Methylheptane | 0.95 |
| 4-Methylheptane | 0.98 |
| 2-Methyloctane | 0.85 |

\*, NMSE values for 2-methylhexane, 2-methyloctane and 2-methylnonane cannot be calculated
290 because the model-related parameter values for these hydrocarbons are not available from our laboratory experiments.

**Table S4:** Model parameters and their estimated values obtained from fitting data to the solutions of the systems of equation (3).

| Parameter * | Value | 95% C.I. | Unit |
|---|---|---|---|
| $B(0)$ | 0.0004 | 0.0001-0.0138 | mmol C |
| $K_f$ | 0.3 | 0.3 | mmol |
| $N_T$ | 327.6 | 327.1 | mmol |
| $K_{g_{C_5}}$ | 56.3 | 16.2-96.4 | mmol |
| $K_{g_{C_6}}$ | 430.3 | 366.1-494.5 | mmol |
| $K_{g_{C_7}}$ | 270.7 | 238.9-302.5 | mmol |
| $K_{g_{C_8}}$ | 90.1 | 69.3-110.9 | mmol |
| $K_{g_{C_9}}$ | 0.9 | 0.71-1 | mmol |
| $K_{g_{C_{10}}}$ | 12.0 | 10.2-13.9 | mmol |
| $K_{g_{toluene}}$ | 4.5 | 4.1-4.8 | mmol |
| $K_{g_{m,p-Xylenes}}$ | 85.1 | 76.9-93.2 | mmol |
| $K_{g_{o-Xylenes}}$ | 17.5 | 14.2-20.8 | mmol |
| $K_{g_{2-MC_6}}$ § | 144.6 | 102.7-186.5 | mmol |
| $K_{g_{3-MC_6}}$ | 144.6 | 102.7-186.5 | mmol |
| $K_{g_{2-MC_7}}$ | 320.4 | 183.8-457.1 | mmol |
| $K_{g_{4-MC_7}}$ | 170.3 | 121.0-219.7 | mmol |
| $K_{g_{2-MC_8}}$ | 335.9 | 179.1-492.9 | mmol |
| $K_{g_{3-MC_8}}$ § | 335.9 | 179.1-492.9 | mmol |
| $K_{g_{2-MC_9}}$ § | 335.9 | 179.1-492.9 | mmol |
| $K_{g_{2-MC_5}}$ | 165.9 | 130.2-201.7 | mmol |
| $C_5 - lag$ | 200 | 200 | days |
| $C_6 - \text{lag}$ | 26 | 26 | days |
| $C_7 - \text{lag}$ | 60 | 40-80 | days |
| $C_8 - \text{lag}$ | 60 | 60 | days |
| $C_9 - \text{lag}$ | 70 | 70 | days |
| $C_{10} - \text{lag}$ | 5 | 5 | days |
| $Toluene - lag$ | 30 | 30 | days |
| $m - \text{and } p - Xylenes - lag$ | 70 | 70 | days |
| $o - Xylenes - lag$ | 60 | 60 | days |
| $2 - \text{MC}_6 - \text{lag}$ § | 25 | 25 | days |
| $3 - \text{MC}_6 - \text{lag}$ | 25 | 25 | days |
| $2 - \text{MC}_7 - \text{lag}$ | 25 | 25 | days |
| $4 - \text{MC}_7 - \text{lag}$ | 25 | 25 | days |
| $2-MC_8-lag$ | 25 | 25 | days |
| $3-MC_8-lag$ § | 25 | 25 | days |
| $2-MC_9-lag$ § | 25 | 25 | days |
| $2-MC_5-lag$ | 23 | 23 | days |

295

* $K_f$ represents the nitrogen-dependent half-saturation constant for microbial growth; $N_T$ is the total nitrogen available in the system; $K_{g_{C_5}}, K_{g_{C_6}}, K_{g_{C_7}}, K_{g_{C_8}}, K_{g_{C_9}}, K_{g_{C_{10}}}, K_{g_{toluene}},$ $K_{g_{o-Xylenes}},$

$K_{g_{m,p-Xylenes}}, K_{g_{3-MC_6}}, K_{g_{2-MC_7}}, K_{g_{4-MC_7}}, K_{g_{2-MC_8}}, K_{g_{3-MC_8}}, K_{g_{2-MC_8}}, K_{g_{2-MC_9}},$

300 $K_{g_{2-MC_5}}$ respectively represent the half-saturation constants for microbial growth on $C_5$-, $C_6$-, $C_7$-, $C_8$-, $C_9$-, $C_{10}$- $n$-alkanes, toluene, o-xylene, m- plus p-xylene, 2-methylhexane-, 3-methylhexane-, 2-methylheptane-, 4-methylheptane-, 2-methyloctane-, 3-methyloctane-, 2-methylnonane- and , 2-methylpentane-. $Z$-lag denotes a lag period of $Z$, where Z is one of $C_5$, $C_6$, $C_7$, $C_8$, $C_9$, $C_{10}$, toluene, $o$-xylene, $m$- plus $p$-xylene, 2-methylhexane-, 3-methylhexane, 2-

305 methylheptane, 4-methylheptane, 2-methyloctane 3-methyloctane, 2-methylnonane or 2-methylpentane.

§ The values of model parameters $K_g$ and lag for 2-$MC_6$, 3-$MC_8$ and 2-$MC_9$ were not available from empirical studies and are assumed to be the same as those for 3-$MC_6$, 2-$MC_8$ and 2-$MC_8$,

310 respectively, based on their similar molecular weights.

**Table S5**: Estimated zero-and first-order model parameter values for labile diluent hydrocarbons
315 not reported by Siddique et al. (2008).

| Hydrocarbon | Lag phase (d) | Zero-order parameter (mmole d$^{-1}$) | First-order parameter (d$^{-1}$) |
|---|---|---|---|
| *n*-Pentane | 294 | 0.0008576 | 0.01117 |
| *n*-Nonane | 77 | 2.664e-05 | 0.01276 |
| 2-Methylpentane | 600 | 0.0002281 | 0.003501 |
| 3-Methylhexane | 455 | 0.0001816 | 0.003849 |
| 2-Methylheptane | 845 | 0.00023 | 0.005258 |
| 4-Methylheptane | 665 | 0.0001936 | 0.005663 |
| 2-Methyloctane | 665 | 0.0001772 | 0.0006584 |

320    **Table S6.** Calculation of mass balance of diluent entering OSTP in 2016 and 2017. These values are used in Table S8 calculations.

|  | Syncrude MLSB | | CNRL Horizon | | CNUL MRM | |
|---|---|---|---|---|---|---|
|  | 2016 | 2017 | 2016 | 2017 | 2016 | 2017 |
| Reported mass of diluent lost to fresh tailings before deposition in OSTP (t) [a] | 57,336 | 43,032 | 24,722 | 35,295 | 28,558 | 32,494 |
| Estimated mass of diluent lost from OSTP by volatilization (t) [b] | (-17,201) | (-12,910) | (-7,416) | (-10,589) | (-11,423) | (-12,998) |
| Calculated net mass of diluent remaining in OSTP (t) | 40,135 | 30,122 | 17,305 | 24,706 | 17,135 | 19,496 |

a, Data retrieved from Alberta Energy Regulator ST 39 report (AER, 2018) and calculated using
325    the reported volume of diluent loss ($m^3$) and multiplying by the respective densities of diluents
(Syncrude naphtha, 0.76 t $m^{-3}$; CNRL naphtha, 0.73 t $m^{-3}$; and CNUL paraffinic solvent, 0.65 t
$m^{-3}$ (Burkus et al., 2014).
b, A factor of 0.7 (i.e., 30% volatilization) was used for Syncrude and CNRL naphtha diluents
and a factor of 0.6 (i.e., 40% volatilization) was used for CNUL paraffinic diluent to calculate
330    the mass of diluent volatilized from OSTP per Burkus et al. (2014) .

**Table S7.** Concentrations of 18 labile hydrocarbons in diluents and calculated masses of labile diluent hydrocarbons present in tailings entering OSTP in 2016 and 2017. Values are used in Table S8 calculations.

| Labile hydrocarbon | Syncrude MSLB | | | CNRL Horizon | | | CNUL MRM | | |
|---|---|---|---|---|---|---|---|---|---|
| | % of naphtha diluent [a] | mass in OSTP (t) 2016 [b] | mass in OSTP (t) 2017 [b] | % of naphtha diluent [a] | mass in OSTP (t) 2016 [b] | mass in OSTP (t) 2017 [b] | % of paraffinic diluent [a] | mass in OSTP (t) 2016 [b] | mass in OSTP (t) 2017 [b] |
| Toluene | 6.11 | 2452 | 1840 | 0 | 0 | 0 | 0 | 0 | 0 |
| $m$-, $p$-Xylene | 4.64 | 1862 | 1398 | 0 | 0 | 0 | 0 | 0 | 0 |
| $o$-Xylene | 1.78 | 714 | 536 | 0 | 0 | 0 | 0 | 0 | 0 |
| $n$-$C_5$ | 0 | 0 | 0 | 0 | 0 | 0 | 24.00 | 4112 | 4679 |
| $n$-$C_6$ | 0.60 | 241 | 181 | 3.85 | 666 | 951 | 11.26 | 1929 | 2195 |
| $n$-$C_7$ | 4.50 | 1806 | 1356 | 9.35 | 1618 | 2310 | 0 | 0 | 0 |
| $n$-$C_8$ | 6.05 | 2428 | 1822 | 4.65 | 805 | 1149 | 0 | 0 | 0 |
| $n$-$C_9$ | 1.99 | 799 | 599 | 1.70 | 294 | 420 | 0 | 0 | 0 |
| $n$-$C_{10}$ | 0.31 | 126 | 94 | 1.65 | 286 | 408 | 0 | 0 | 0 |
| 2-$MC_5$ | 0 | 0 | 0 | 1.25 | 216 | 309 | 23.50 | 4027 | 4582 |
| 2-$MC_6$ | 1.30 | 522 | 392 | 5.30 | 917 | 1309 | 0 | 0 | 0 |
| 3-$MC_6$ | 1.51 | 607 | 456 | 5.05 | 874 | 1248 | 0 | 0 | 0 |
| 2-$MC_7$ | 4.92 | 1976 | 1483 | 3.85 | 666 | 951 | 0 | 0 | 0 |
| 4-$MC_7$ | 1.86 | 747 | 561 | 1.25 | 216 | 309 | 0 | 0 | 0 |
| 2-$MC_8$ | 1.16 | 465 | 349 | 1.00 | 173 | 247 | 0 | 0 | 0 |
| 3-$MC_8$ | 1.55 | 623 | 467 | 0.55 | 95 | 136 | 0 | 0 | 0 |
| 2-$MC_9$ | 0.31 | 124 | 93 | 2.90 | 502 | 717 | 0 | 0 | 0 |
| % of diluent considered labile | 39 | | | 42 | | | 59 | | |
| Total mass of labile hydrocarbon entering OSTP (t) | | 15492 | 11627 | | 7329 | 10463 | | 10068 | 11456 |

[a] The concentrations of individual hydrocarbons in Syncrude and CNRL naphtha diluents were calculated using PONAU analysis reported by (Siddique et al., 2007) and (Mohamad Shahimin, and Siddique, 2017b), respectively, and the concentrations of individual hydrocarbons in CNUL paraffinic diluent were calculated using the PONAU analysis reported by (Mohamad Shahimin and Siddique, 2017a).

[b] The data were retrieved from Alberta Energy Regulator report ST 39 (AER, 2018)

**Table S8:** Contribution of individual labile diluent hydrocarbons to the maximum theoretical cumulative yield of $CH_4$ from OSTPs in 2016 and 2017, based on masses calculated in Tables S5 and S6). Methane yield was calculated using equation (4) in main text, per Symons and Buswell (1933) as implemented by Roberts (2002).

| Labile hydrocarbon | Calculated theoretical methane production (moles x 10[6]) | | | | | |
|---|---|---|---|---|---|---|
| | Syncrude MLSB | CNRL Horizon | CNUL MRM | Syncrude MLSB | CNRL Horizon | CNUL MRM |
| | 2016 | | | 2017 | | |
| Toluene | 120 | 0 | 0 | 90 | 0 | 0 |
| $m$-, $p$-Xylene | 92 | 0 | 0 | 69 | 0 | 0 |
| $o$-Xylene | 35 | 0 | 0 | 27 | 0 | 0 |
| $n$-$C_5$ | 0 | 0 | 228 | | 0 | 259 |
| $n$-$C_6$ | 13 | 37 | 106 | 10 | 52 | 121 |
| $n$-$C_7$ | 99 | 89 | 0 | 74 | 127 | 0 |
| $n$-$C_8$ | 133 | 44 | 0 | 100 | 63 | 0 |
| $n$-$C_9$ | 44 | 16 | 0 | 33 | 23 | 0 |
| $n$-$C_{10}$ | 7 | 16 | 0 | 5 | 22 | 0 |
| 2-$MC_5$ | 0 | 12 | 222 | 0 | 17 | 253 |
| 2-$MC_6$ | 29 | 50 | 0 | 21 | 72 | 0 |
| 3-$MC_6$ | 33 | 48 | 0 | 25 | 69 | 0 |
| 2-$MC_7$ | 108 | 36 | 0 | 81 | 52 | 0 |
| 4-$MC_7$ | 41 | 12 | 0 | 31 | 17 | 0 |
| 2-$MC_8$ | 25 | 9 | 0 | 19 | 14 | 0 |
| 3-$MC_8$ | 34 | 5 | 0 | 25 | 7 | 0 |
| 2-$MC_9$ | 7 | 27 | 0 | 5 | 39 | 0 |
| Total theoretical methane (moles x 10[6]) [a] | 820 | 401 | 556 | 615 | 574 | 633 |
| Microbial hydrocarbon conversion to methane (moles x 10[6]) [b] | 656 | 321 | 445 | 492 | 459 | 506 |
| Total methane emissions from ponds (moles x 10[6]) [c] | 1191 | 336 | 2634 | 991 | 599 | 1051 |
| Contribution of diluent hydrocarbons to total methane emissions from ponds (%) | 55 | 95 | 17 | 50 | 77 | 48 |

[a] The masses of individual hydrocarbons from Table S6 were converted into moles using the respective molecular weights and then Symons and Buswell equation (per Roberts, 2002) was used to calculate theoretical maximum methane production from individual hydrocarbons.

[b] A factor of 0.8 determined during our hydrocarbon biodegradation studies (Siddique et al., 2007, 2006) was used to calculate the efficiency of microbial conversion of hydrocarbons to methane; i.e., $r_i$

[c] $CH_4$ emission data (unpublished data, Government of Alberta) were converted into moles for comparison. The Government of Alberta data includes $CH_4$ emissions from all units. We considered only those units that had been receiving froth treatment tailings (solvent containing stream) for the most recent two or three years. Therefore, for comparison, the bubbling zone of Syncrude MLSB, the entire CNRL Horizon pond and Cells 1-3 of CNUL receiving diluent containing streams were used for field emissions data.

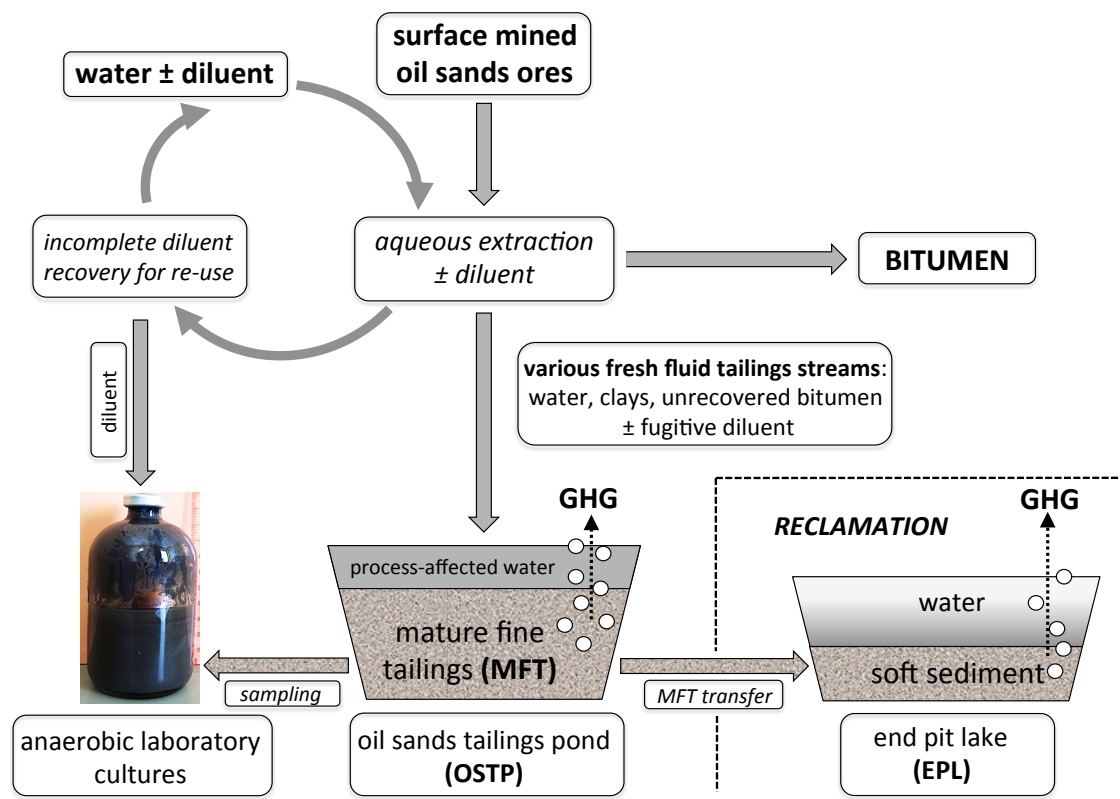

**Figure S1.** Simplified schematic of aqueous bitumen extraction from surface-mined oil sands, with subsequent retention of tailings in oil sands tailings ponds (OSTP) and reclamation in end pit lakes (EPL) (reviewed Foght et al., 2017). Biogenic gases in tailings (1) may escape to the atmosphere from shallow sediments via ebullition as greenhouse gas (GHG) emissions during retention or from deeper sediments when physically disturbed (e.g., by mechanical transfer), or (2) may be trapped as temporary or permanent gas voids (Guo, 2009) in dense sediments as latent GHG emissions, or (3) may be immobilized and transformed via geochemical interactions with clay minerals and pore water (Siddique et al., 2014).

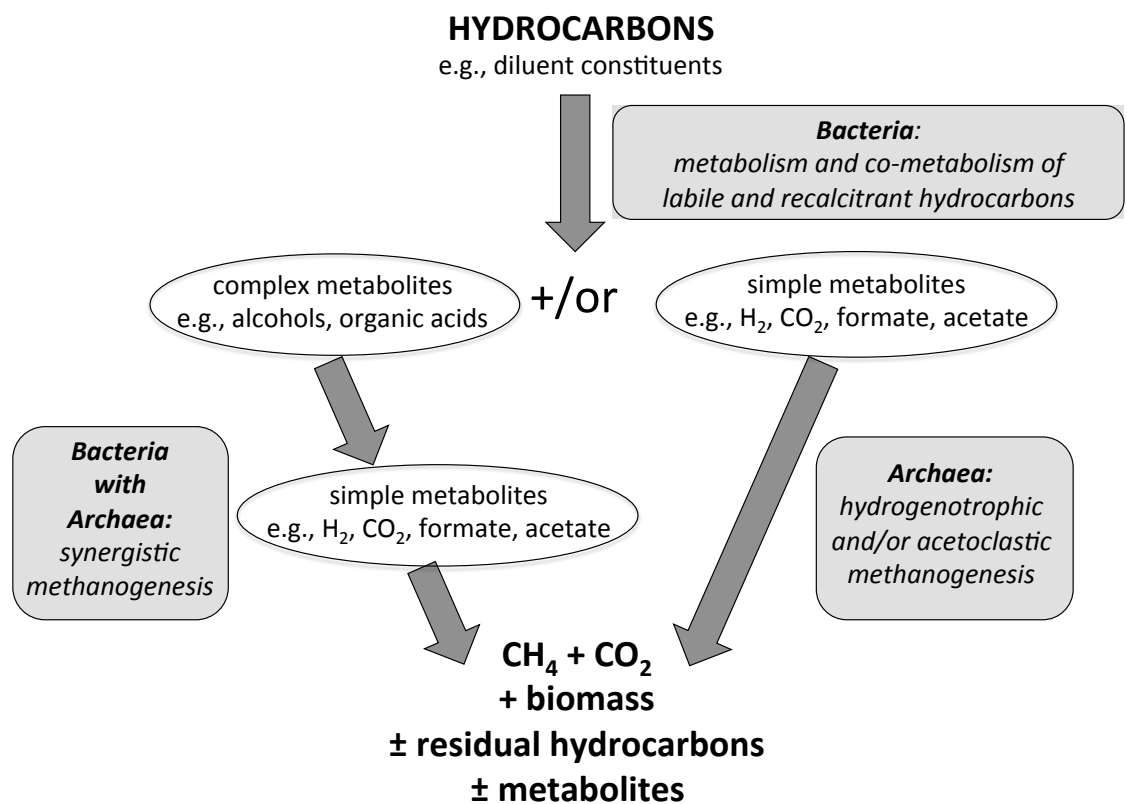

**Figure S2.** Simplified biochemical flowchart for methanogenic biodegradation of hydrocarbons.
Metabolic processes carried out by Bacteria or Archaea alone or by synergistic consortia are
indicated in italics. If sulfate is present in sufficient concentrations (e.g., via addition of gypsum
[CaSO$_4$•2H$_2$O] in some oil sands tailing processes; Foght et al., 2017), anaerobic biodegradation
may still proceed but will be skewed toward accumulation of metabolites plus CO$_2$ and biomass,
with minimal CH$_4$ production. The ultimate end products include GHG, biomass, non-degradable
hydrocarbons and dead-end metabolites, e.g., from partial oxidation of recalcitrant hydrocarbons.

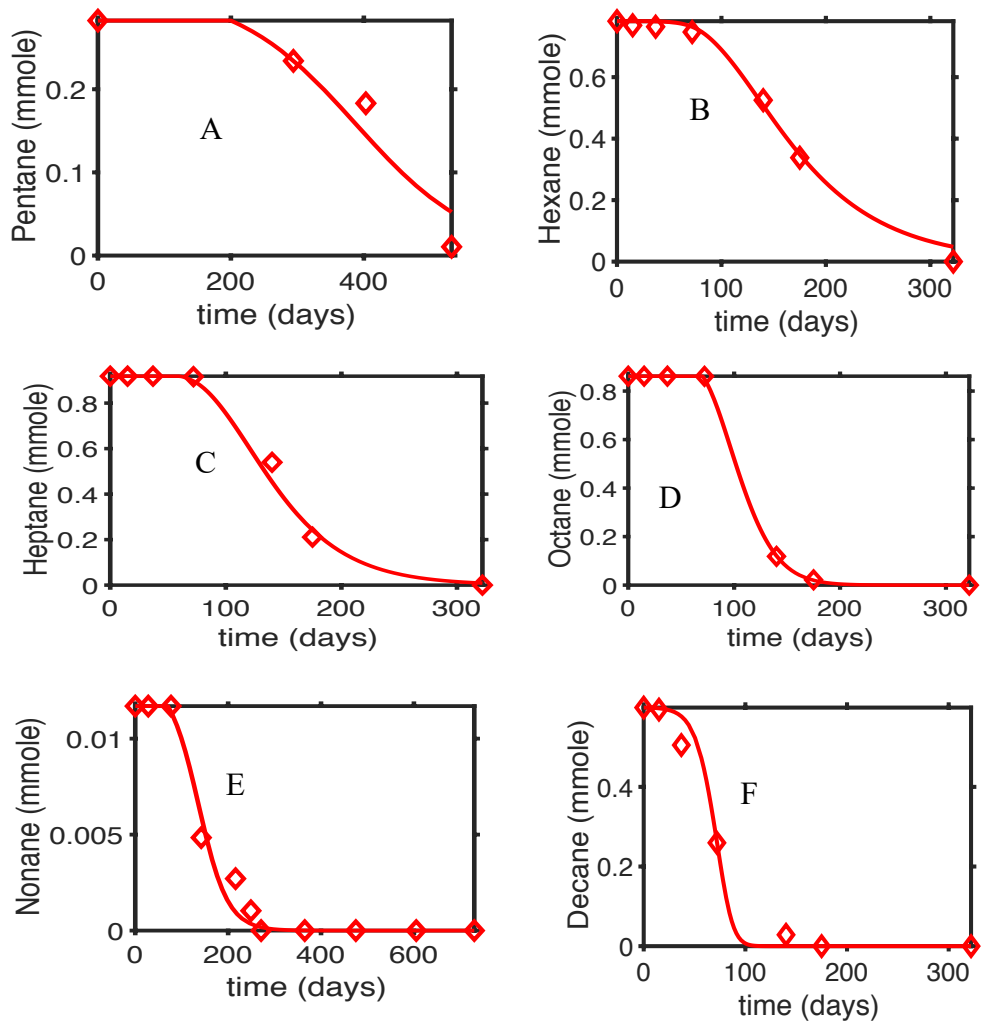

**Figure S3.** System of equations (2) fit to measured *n*-alkane biodegradation values for laboratory
390   cultures. Symbols denote measured values and lines represent best fits to the data. Panels A, B,
C, D, E and F show results for *n*-pentane, *n*-hexane, *n*-heptane, *n*-octane, *n*-nonane, and *n*-
decane, respectively.

395

400

405

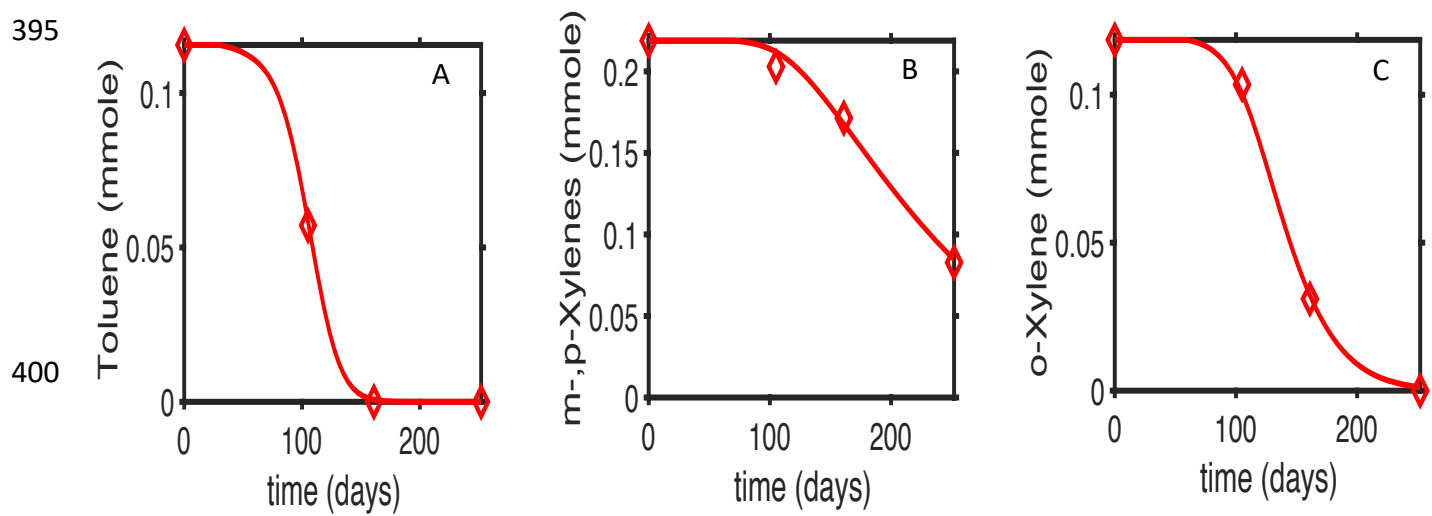

**Figure S4.** System (2) fit to measured biodegradable monoaromatic compound data for laboratory cultures. Diamond symbols denote measured values and solid lines represent fitted values. Panels A, B and C respectively show results for toluene, *m*- plus *p*-xylene, and *o*-xylene.

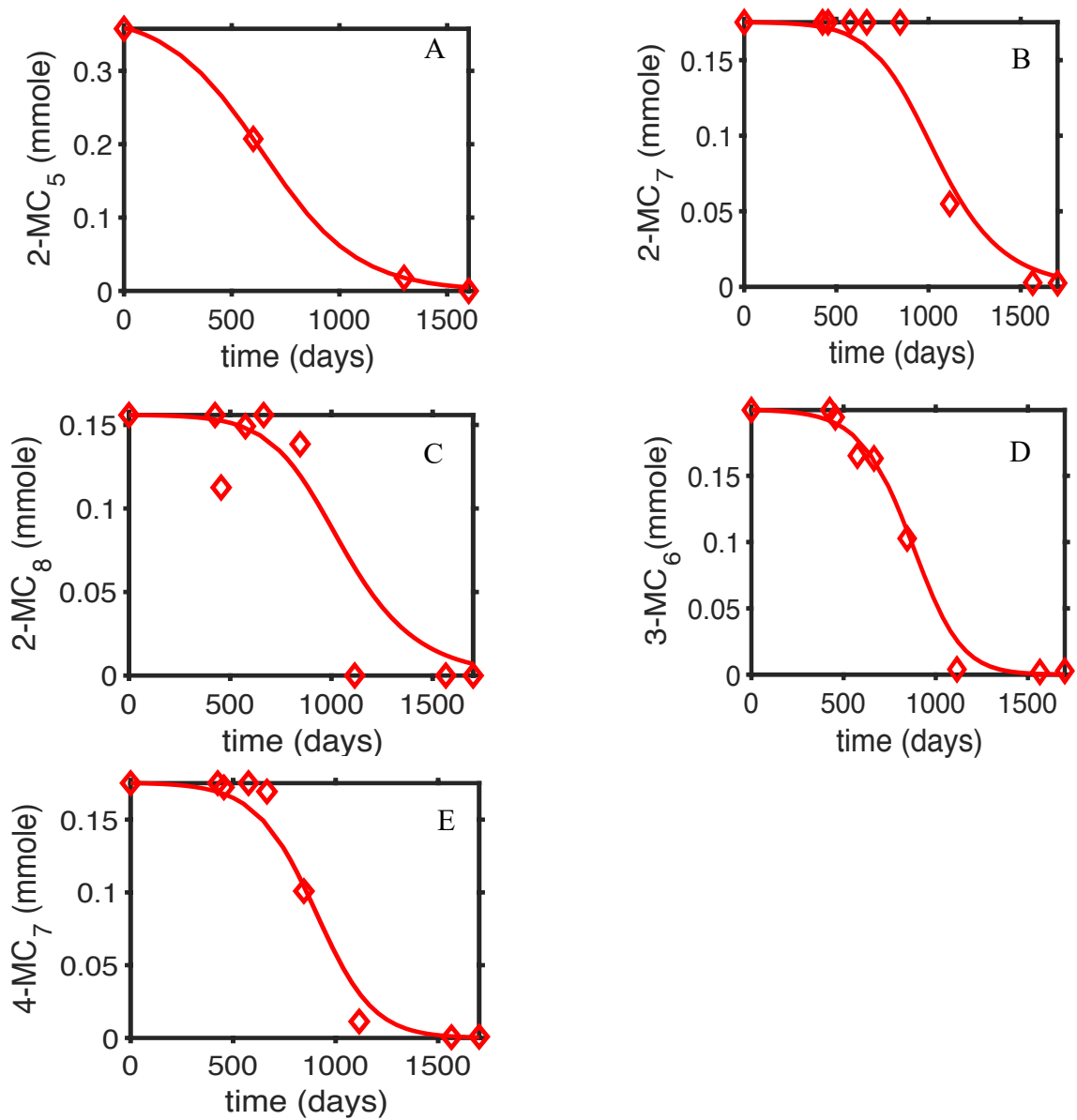

410

**Figure S5.** System (2) fit to *iso*-alkane biodegradation measurements for laboratory cultures.
Solid lines represent fitted values and diamonds denote measured values. Panels A, B, C, D and
415    E show results for 2-methylheptane, 2-methyloctane, 2-methylpentane, 3-methylhexane and 4-methylheptane, respectively.

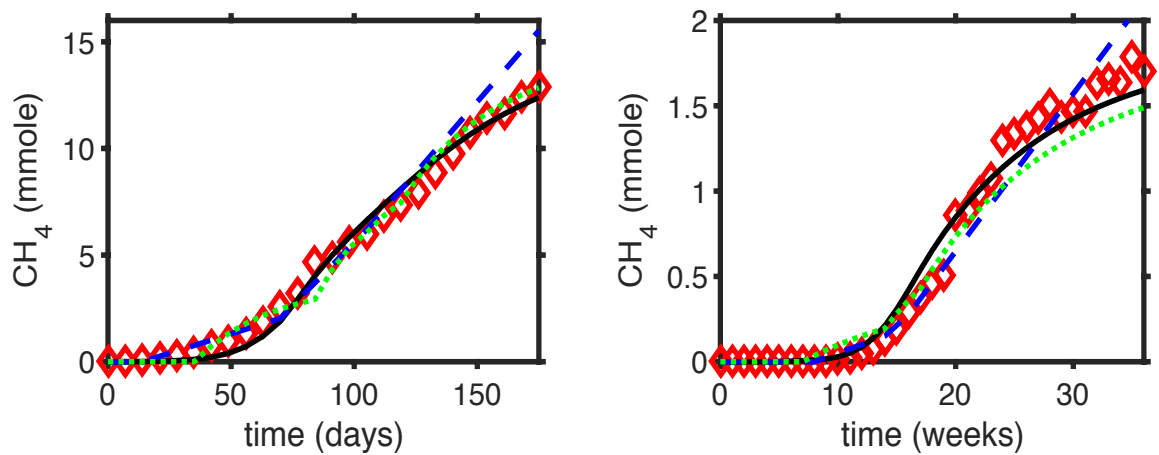

**Figure S6:** Comparison of stoichiometric model predictions of methane production from
laboratory cultures of Syncrude MFT incubated with mixtures of either *n*-alkane ($C_6$, $C_7$, $C_8$ and
$C_{10}$) or monoaromatic (toluene, *o*-, *m*- and *p*-xylenes) components of naphtha diluent (left and
right panels, respectively). Measured methane values, from laboratory experiments independent
of those used to develop the model, are shown by diamond symbols. Solid black lines represent
the stoichiometric model prediction; broken blue lines and dotted green lines respectively
represent predictions made by using the previous zero-order and first-order models (Siddique et
al., 2008).

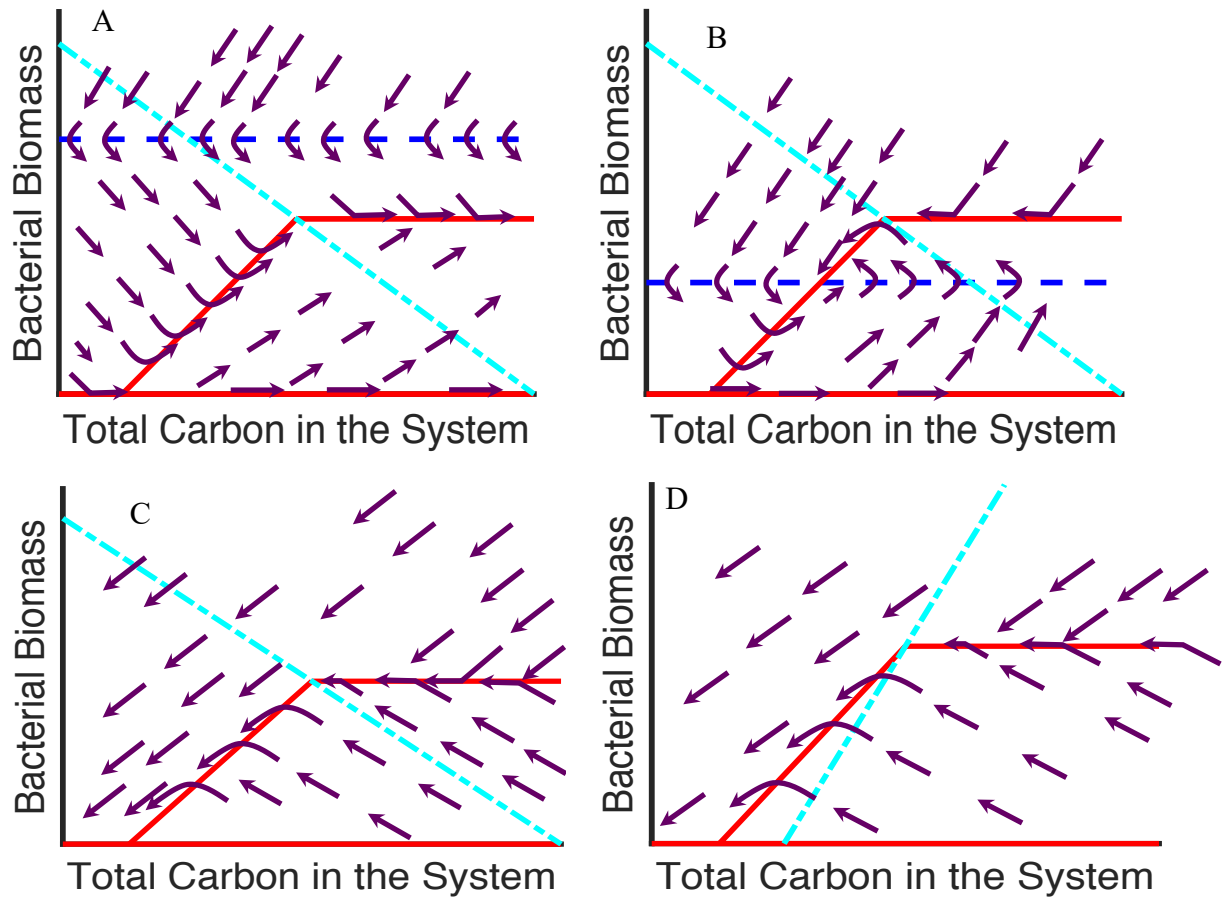

430

**Figure S7:** Phase plane analysis of solution states for microbial biomass and total carbon content in OSTP (Panels A and B, where $C^{in} > 0$) or EPL (Panels C and D, where $C_T^{in} = 0$) under different assumed initial conditions of $C_T{}^{in}$ and ratio of the nitrogen carrying capacity to carbon carrying capacity ($k_f : k_g$). In Panel A: $C_T^{in} > \left(N_T - \frac{dk_f}{\mu}\right)\frac{d(1-r)}{\theta r}$ and $k_f : k_g < \theta r$. In Panel B,

435 $C_T^{in} < \left(N_T - \frac{dk_f}{\mu}\right)\frac{d(1-r)}{\theta r}$ and $k_f : k_g < \theta r$. In Panel C: $k_f : k_g < \theta r$. In Panel D: $k_f : k_g > \theta r$.

Solid red lines are nullclines for total biomass, broken blue lines are nullclines for total carbon content and broken light blue lines indicate where $B = \left(N_T - \frac{\left(C_T + \frac{B}{r}\right)k_f}{k_g}\right)\left(\frac{k_g r}{\theta k_g r - k_f}\right)$, to the left of which nitrogen is limiting and to the right of which carbon is limiting. The slope of this line is determined by the ratio: $k_f : k_g$. Purple directional arrows account for time.

440

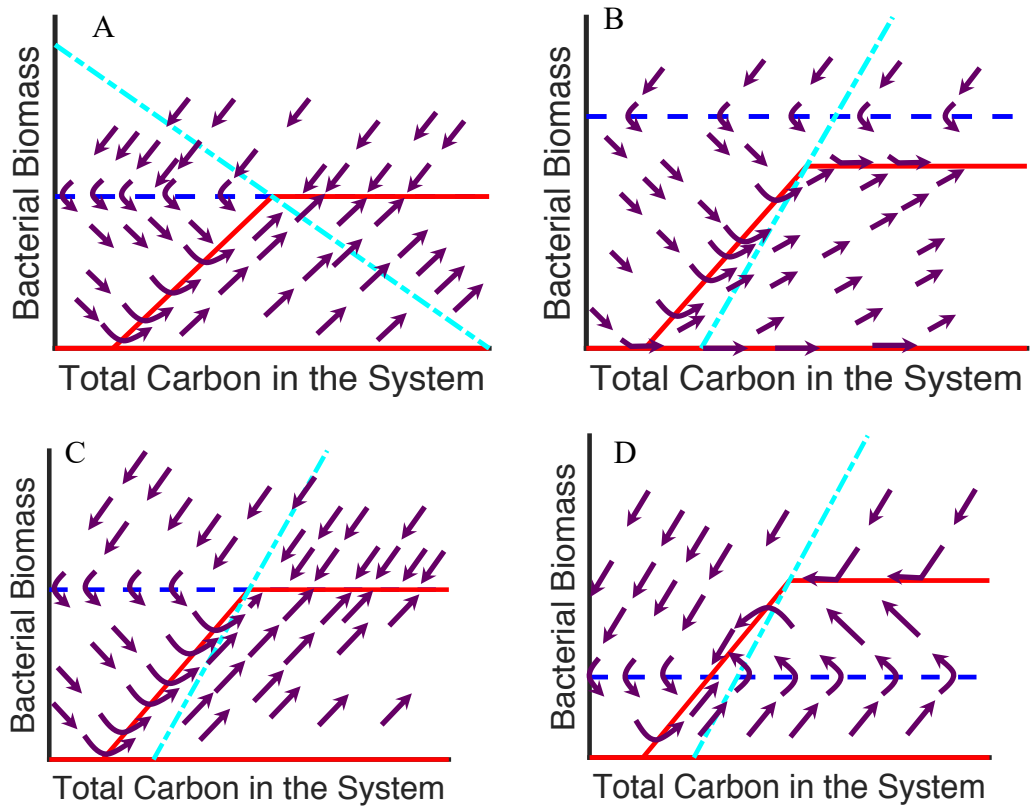

445 **Figure S8:** Phase plane analysis of solution states for microbial biomass and total carbon content in OSTP (where $C^{in} > 0$) under different assumed initial conditions of $C_T^{in}$ and ratio of the nitrogen carrying capacity to carbon carrying capacity ($k_f : k_g$). In Panel A: $C_T^{in} = \left(N_T - \frac{dk_f}{\mu}\right)\frac{d(1-r)}{\theta r}$ and $k_f : k_g < \theta r$. In Panel B, $C_T^{in} > \left(N_T - \frac{dk_f}{\mu}\right)\frac{d(1-r)}{\theta r}$ and $k_f : k_g > \theta r$. In Panel C: $C_T^{in} = \left(N_T - \frac{dk_f}{\mu}\right)\frac{d(1-r)}{\theta r}$ and $k_f : k_g > \theta r$. In Panel D: $C_T^{in} < \left(N_T - \frac{dk_f}{\mu}\right)\frac{d(1-r)}{\theta r}$ and $k_f : k_g > \theta r$. Solid red lines are nullclines for total biomass, broken blue lines are nullclines for total carbon content and broken light blue lines indicate where the line $B = \left(N_T - \frac{\left(C_T + \frac{B}{r}\right)k_f}{k_g}\right)\left(\frac{k_g r}{\theta k_g r - k_f}\right)$ to the left of which nitrogen is limiting and to the right of which carbon is limiting. The slope of this line is determined by the ratio: $k_f : k_g$. Purple directional arrows account for time.

455